\documentclass[twocolumn,times]{aastex63} % 

\usepackage{amsmath}
\usepackage{amsfonts}
\usepackage{booktabs}
\usepackage{multirow}
\usepackage{txfonts}

\newcommand{\diff}{\textrm{d}}
\newcommand{\E}{\text{E}}
\newcommand{\ignore}[1]{}

\newcommand{\myvec}[1]{\boldsymbol{\mathbf{#1}}}
\newcommand{\mymatrix}[1]{\boldsymbol{\mathbf{#1}}}

\newcommand{\pder}[2]{\frac{\partial#1}{\partial#2}}

\received{April 12, 2021}
\revised{July 21, 2021}
\accepted{July 28, 2021}
\submitjournal{AJ}

\shorttitle{A novel approach to asteroid impact monitoring and hazard assessment}
\shortauthors{Roa et~al.}

\graphicspath{{./}{figures/}}

\begin{document}

\title{A novel approach to asteroid impact monitoring and hazard assessment}

\correspondingauthor{Javier Roa\\ \copyright 2021. California Institute of Technology. Government sponsorship acknowledged.}
\email{javier.roa@jpl.nasa.gov}

\author[0000-0002-0810-1549]{Javier Roa}
\affiliation{Jet Propulsion Laboratory, California Institute of Technology\\
4800 Oak Grove Dr \\
Pasadena, CA 91109, USA}

\author[0000-0003-0774-884X]{Davide Farnocchia}

\affiliation{Jet Propulsion Laboratory, California Institute of Technology\\
4800 Oak Grove Dr \\
Pasadena, CA 91109, USA}

\author[0000-0003-3240-6497]{Steven R. Chesley}

\affiliation{Jet Propulsion Laboratory, California Institute of Technology\\
4800 Oak Grove Dr \\
Pasadena, CA 91109, USA}

\begin{abstract}
Orbit-determination programs find the orbit solution that best fits a set of observations by minimizing the RMS of the residuals of the fit. For near-Earth asteroids, the uncertainty of the orbit solution may be compatible with trajectories that impact Earth. This paper shows how incorporating the impact condition as an observation in the orbit-determination process results in a robust technique for finding the regions in parameter space leading to impacts. The impact pseudo-observation residuals are the $b$-plane coordinates at the time of close approach and the uncertainty is set to a fraction of the Earth radius. The extended orbit-determination filter converges naturally to an impacting solution if allowed by the observations. The uncertainty of the resulting orbit provides an excellent geometric representation of the virtual impactor. As a result, the impact probability can be efficiently estimated by exploring this region in parameter space using importance sampling. The proposed technique can systematically handle a large number of estimated parameters, account for nongravitational forces, deal with nonlinearities, and correct for non-Gaussian initial uncertainty distributions. The algorithm has been implemented into a new impact monitoring system at JPL called Sentry-II, which is undergoing extensive testing. The main advantages of Sentry-II over JPL's currently operating impact monitoring system Sentry are that Sentry-II can systematically process orbits perturbed by nongravitational forces and that it is generally more robust when dealing with pathological cases. The runtimes and completeness of both systems are comparable, with the impact probability of Sentry-II for 99\% completeness being $3\times10^{-7}$.
\end{abstract}


\section{Introduction}
%Asteroid orbits are determined by finding the solution that best fits a set of observations. In practice, the vector of parameters that uniquely defines an orbit is treated as a random vector to account for the multiple sources of errors that limit the accuracy of the observations and the dynamical model. Its expected value is the most likely orbit of the asteroid and the uncertainty in the knowledge of the orbit is represented by a probability density function \citep{milani2010theory,tapley2004statistical,wiesel2003modern}.

The set of parameters defining an asteroid's orbit is usually modeled as a random vector following a multivariate Gaussian distribution \citep{milani2010theory}. The mean represents the nominal orbit and the covariance characterizes the uncertainty in the knowledge of the orbit. The orbital uncertainty distribution outlines the region in parameter space where the actual orbit of the asteroid may be found within a certain confidence level, with orbits closer to the nominal being more likely to represent the true orbit. Some of these orbits may impact Earth. It is the role of impact monitoring to identify such impacting orbits and to estimate the probability of the true orbit being an impactor.

Monte Carlo (MC) simulation is the simplest and most robust technique for estimating impact probabilities because it introduces no assumptions or simplifications when propagating the uncertainty distribution over time \citep{chodas1999orbit,chodas1999predicting,farnocchia2013yarkovsky}. Multiple samples are drawn from the orbital uncertainty distribution at the initial conditions, propagated over time using the fully nonlinear equations of motion, and the distribution of impacts is treated as a binomial distribution. However, MC sampling is computationally expensive and impractical for estimating small probabilities due to the large number of samples required to obtain an estimate with reasonable variance \citep{farnocchia2015orbits}. \citet{milani1999asteroid} introduced the concept of the line of variations (LOV) to reduce the dimension of the space of parameters to be explored, resulting in a more efficient sampling technique. The LOV is a one-dimensional subspace that captures the weak direction---and therefore most of the arising nonlinearities---of the uncertainty region when sampled and propagated over time  \citep{milani2002asteroid,milani2005nonlinear,milani2005multiple,farnocchia2015orbits}. Since one can interpolate between virtual asteroids distributed along the LOV to find impacting solutions, this technique provides a so far unmatched ability to detect small impact probability events in the nonlinear regime with little computational effort and to identify distinct dynamical pathways. For this reason, the operational impact-monitoring systems Sentry\footnote{https://cneos.jpl.nasa.gov/sentry/} \citep[operated by JPL, see][]{chamberlin2001sentry} and CLOMON2\footnote{https://newton.spacedys.com/neodys2/index.php?pc=4.1} (operated by the University of Pisa) are both based on this technique \citep{milani2005multiple}.

Impact monitoring systems based on the LOV have been extremely successful over the last 20~years. However, some assumptions limit the applicability of the method in certain cases. For instance, sampling the LOV requires choosing a distance metric for the orbital elements, which is to some extent arbitrary \citep{milani2005multiple}. Choosing the right metric is expected to capture the dynamical dispersion of the uncertainty region accurately. This is not always assured and defining an adequate metric has been particularly challenging when nongravitational effects need to be accounted for, as in the case of asteroid (101955) Bennu \citep{chesley2014orbit} and (29075) 1950~DA \citep{farnocchia2014assessment}. \citet{spoto2014nongravitational} addressed this issue by identifying a scattering encounter, finding the weak direction on the $b$-plane \citep{farnocchia2019planetary}, and then mapping it back to the space of initial conditions and nongravitational parameters. Given these limitations, MC sampling is still the preferred option for estimating the impact probability of asteroids subject to nongravitational accelerations \citep{farnocchia2013yarkovsky,farnocchia2014assessment,farnocchia2017mass,reddy2019near}. Other known issues of the LOV include its possible indeterminacy for orbits determined from very short arcs, and certain pathological cases where the LOV may have such a complicated geometry on the $b$-plane that interpolating between virtual asteroids may not identify all possible impactors. \citet{del2020improving} improved the performance of the LOV method in these scenarios using an adaptive densification of the LOV. A dedicated discussion of the limitations of the LOV technique can be found in \citet{farnocchia2015orbits}.

With the perspective of the number of known near-Earth asteroids (NEAs) increasing substantially in the coming years when new surveys such as LSST become operational \citep{ivezic2019lsst}, there is a renewed interest in finding robust impact monitoring techniques that overcome these limitations by being able to systematically handle estimated parameters while being as efficient as the LOV method. \citet{del2020manifold} improved the robustness of the LOV method when dealing with short arcs, and \citet{losacco2018differential} and \citet{roa2019multilayer} presented alternative techniques for identifying the subdomains of the uncertainty region where impact solutions are likely to be found, effectively reducing the size of the region to be explored. \citet{losacco2018differential} rely on differential algebra combined with importance sampling, while \citet{roa2019multilayer} present a sequential MC strategy. These techniques still require some manual intervention to efficiently explore the space of parameters, which limits their implementation into fully automatic systems.

This paper introduces a new method to robustly detect and characterize impact trajectories based on orbit determination techniques. We implemented this method into a new automatic impact-monitoring system, called Sentry-II, that is currently undergoing extensive testing. A slight modification to the conventional least-squares filter makes the filter converge to an impact trajectory as long as it is compatible with the available observations. After summarizing some key concepts of orbit determination in Section~\ref{Sec:orbit_determination}, Section~\ref{Sec:impact_monitoring} explains how to incorporate the impact pseudo-observation into the orbit-determination filter to find impact trajectories and how to estimate the impact probability. Section~\ref{Sec:discussion} discusses practical aspects of the proposed technique, including how to deal with nonlinear dynamics, how to account for non-Gaussian uncertainty distributions in the probability estimate, and the completeness of the system. Finally, Sections~\ref{Sec:examples} and~\ref{Sec:runtime} assess the performance of the system through multiple examples and comparison with existing techniques.

\section{Orbit determination}\label{Sec:orbit_determination}
Let $\myvec{x}\in\mathbb{R}^n$ denote the vector of parameters to be estimated, which usually reduces to the state vector or a set of orbital elements at a reference epoch. In some cases, additional parameters like the mass of a perturber \citep{baer2017simultaneous} or the parameters in the Yarkovsky model \citep{bottke2006yarkovsky,chesley2015direct,farnocchia2013near,farnocchia2013yarkovsky} can also be estimated to model the orbit of the asteroid.

Given a set of observations of the asteroid along its orbit, the observed values can be compared with predictions obtained by integrating the equations of motion starting from some informed guess of $\myvec{x}$. The difference between the observed values and the predicted values defines the vector of residuals, $\myvec{e}$. Different values of $\myvec{x}$ produce different orbits and residuals. The goal of orbit determination is to find the orbit that best fits the data. That is, the value of $\myvec{x}$ that minimizes the cost function
\begin{equation}\label{Eq:cost_od}
    Q=\myvec{e}^\top\mymatrix{W}\myvec{e},
\end{equation}
where $\mymatrix{W}$ is the weight matrix. In the simplest case where the observation errors are uncorrelated, $\myvec{W}$ becomes diagonal with entries $1/\sigma_i^2$, where $\sigma_i$ is typically the uncertainty associated with the $i$-th observation. 

%  The weight matrix is symmetric and positive definite.

The least-squares filter \citep[see, for example,][]{wiesel2003modern,tapley2004statistical,milani2010theory} resorts to a differential-corrector scheme to solve for the parameters based on
\begin{equation}\label{Eq:od_condition}
    \pder{Q}{\myvec{x}} = 2\myvec{e}^\top\mymatrix{W} \mymatrix{B}= \myvec{0}, \qquad \mymatrix{B} = \pder{\myvec{e}}{\myvec{x}}.
\end{equation}
In this equation, $\mymatrix{B}$ is the design matrix and describes how the residuals change as a function of the parameters $\myvec{x}$. Introducing the information matrix
\begin{equation}
    \mymatrix{C} = \mymatrix{B}^\top\mymatrix{W}\mymatrix{B},
\end{equation}
Eq.~\eqref{Eq:od_condition} can be solved iteratively thanks to the relation
\begin{equation}\label{Eq:diffcorr}
    \myvec{x}_{k+1} - \myvec{x}_{k} = -\mymatrix{C}^{-1}\mymatrix{B}^\top\mymatrix{W}\myvec{e}_k
\end{equation}
until the change in the RMS of the residuals or the correction $||\myvec{x}_{k+1}-\myvec{x}_{k}||$ are smaller than prescribed tolerances. The solution $\bar{\myvec{x}}$ defines the nominal orbit of the asteroid. This approach is equivalent to expanding $\partial Q/\partial\myvec{x}$ in series, assuming that the residuals are small, and truncating the series to first order. Alternative methods for sequential estimation include the square-root information filter \citep{bierman1977factorization} and the Kalman filter \citep{kalman1960new}.

In the linear regime and assuming that the observation errors are normally distributed, $\myvec{x}$ is normally distributed with mean $\bar{\myvec{x}}$ and covariance matrix
\begin{equation}\label{Eq:covariance_filter}
    \mymatrix{\Sigma} = \mymatrix{C}^{-1}.
\end{equation}
The probability density function (p.d.f.) associated with the solution is
\begin{equation}\label{Eq:pdf_q}
    q(\myvec{x})=\mathcal{N}(\bar{\myvec{x}},\mymatrix{\Sigma}),
\end{equation}
where $\mathcal{N}$ denotes the normal distribution with mean $\bar{\myvec{x}}$ and covariance $\mymatrix{\Sigma}$. Although in theory $q(\myvec{x})>0$ for any $\myvec{x}\in\mathbb{R}^n$, in practice the parameter space can be reduced to the subdomain
\begin{equation}
	{D}_\sigma = \{ \myvec{x}\in\mathbb{R}^n\; |\; \myvec{x}^\top \mymatrix{\Sigma}^{-1}\myvec{x} < \sigma^2 \}
\end{equation}
given a confidence level $\sigma$.

The orbit can be parameterized using either Cartesian coordinates or a set of elements. Although orbital elements can be better suited for numerical propagation \citep{stiefel1971linear,roa2017regularization}, we choose Cartesian coordinates since they are nonsingular and, for short-arcs, they are more robust for orbit determination and their uncertainty region is better approximated by an ellipsoid \citep{milani2005nonlinear,milani2005multiple}.

 %The set of all possible initial conditions compatible with $q$ is denoted ${D}$, i.e., $\myvec{x}\in{D}$.

\section{Impact monitoring}\label{Sec:impact_monitoring}
The subdomain ${D}_\sigma$ may contain trajectories that impact the Earth when propagated over time. \citet{milani2005nonlinear} introduced the concept of virtual impactor (VI) to refer to a region $F_i$ in parameter space containing trajectories that impact Earth on a certain date and following the same dynamical path. Thus, each $\myvec{x}\in F_i$ is an impact trajectory. The impact probability of a given VI is
\begin{equation}\label{Eq:ip_definition}
    P(F_i) = \int_{F_i} q(\myvec{x})\, \text{d}\myvec{x}.
\end{equation}
Alternatively, the integral expression in Eq.~\eqref{Eq:ip_definition} can be written as
\begin{equation}\label{Eq:ip_definition_If}
    P(F_i) = \int_{{D}_\sigma} I_{F_i}(\myvec{x}) q(\myvec{x})\, \text{d}\myvec{x}
\end{equation}
in terms of the indicator function $I_{F_i}(\myvec{x})$, which is one for $\myvec{x}\in F_i$ and zero otherwise. The uncertainty region of an asteroid may admit multiple disjoint VIs that result in impacts on different dates or even on the same date but following different dynamical paths.

The confidence level $\sigma$ must be chosen so that $D_\sigma$ contains all relevant VIs. For Sentry-II, we set $\sigma=7$, which makes the probability of the actual orbit being outside ${D}_7$ less than $8\times10^{-9}$ when $\myvec{x}\in\mathbb{R}^6$. Impact trajectories found outside this subdomain are discarded as statistically insignificant.

%In this paper, the minimum impact probability that an impact-monitoring system must detect is assumed to be $10^{-7}$. Therefore,

All the VIs producing impacts within a prescribed time span (typically 100~years) must be identified to assess the risk posed by an asteroid. The operation of an automatic impact-monitoring system can be divided into two steps:
\begin{enumerate}
    \item Find all the VIs in the orbital uncertainty distribution $q(\myvec{x})$ within a certain confidence level, $F_i\subset{D}_7$, and with impact probability greater than a prescribed threshold. %This step can be divided in two phases: first, find the dates of all possible close approaches; second, for each close approach, identify and characterize all the disjoint subsets that result in impacts on that date.
    \item Estimate the impact probability of each VI by evaluating the integral of the p.d.f.\ $q(\myvec{x})$ over $F_i$ according to Eq.~\eqref{Eq:ip_definition}.
\end{enumerate}

Additionally, our proposed method characterizes each VI by providing the nominal impacting orbit plus a representation of the region in parameter space leading to impacts along the same dynamical path.

\subsection{Finding virtual impactors\label{Sec:find_VIs_general}}
The process of finding VIs is divided into two steps: first, identifying the close approaches; second, looking for impact trajectories for each close approach.

\subsubsection{Close-approach detection}\label{Sec:ca_detection}
Identifying the dates when close approaches may occur constitutes a preliminary filter that reduces the search space when looking for VIs. In this context, close approach means that the asteroid comes within 0.1~au of the Earth. 

To identify the close-approach dates, a set of $N$ virtual asteroids is sampled from the initial uncertainty distribution $q(\myvec{x})$ and their orbits are propagated 100~years into the future. All Earth close approaches within 0.1~au are tabulated and sorted by close-approach date. Following Sentry, the Sentry-II system samples $N=10^4$ virtual asteroids from $q(\myvec{x})$. The choice of $N$ is driven by the desired impact probability resolution of the impact-monitoring system. Section~\ref{Sec:completeness} assesses the completeness level of the system from a statistical perspective. By sampling the $N$ virtual asteroids from the $n$-dimensional distribution of orbital uncertainty directly, we make no assumptions about how impacting solutions are distributed in parameter space. This $n$-dimensional exploration is an advantage over the LOV method, which depends on the a priori choice of the metric that represents the LOV.

%The completeness level of the proposed algorithm is quantified in Section~\ref{Sec:completeness}.

Next, a sliding-window filter scans the list of close-approach times sequentially and clusters consecutive close approaches as long as their difference is smaller than a time tolerance, which we set to 45~days. This is the maximum time separation between consecutive close approaches in a given cluster, and also the minimum time separation between consecutive clusters. The dispersion of dates within each cluster is an indicator of the uncertainty in the knowledge of that particular close-approach date. When the dispersion is large or when the time separation between close approaches is small (as in the case of temporarily captured objects), clusters corresponding to distinct close approaches may overlap, and the sliding-window filter may not be able to split them. To solve this problem, as we advance sequentially through the list of close-approach dates, we start a new cluster when the same virtual asteroid appears twice in the same cluster even if the time separation with the previous sample is smaller than the tolerance. A similar splitting technique is discussed in \citet{del2019completeness}. The close-approach clustering is usually more efficient for inclined orbits when close approaches occur at the node crossing.

\begin{figure}
    \centering
    \includegraphics[width=\linewidth]{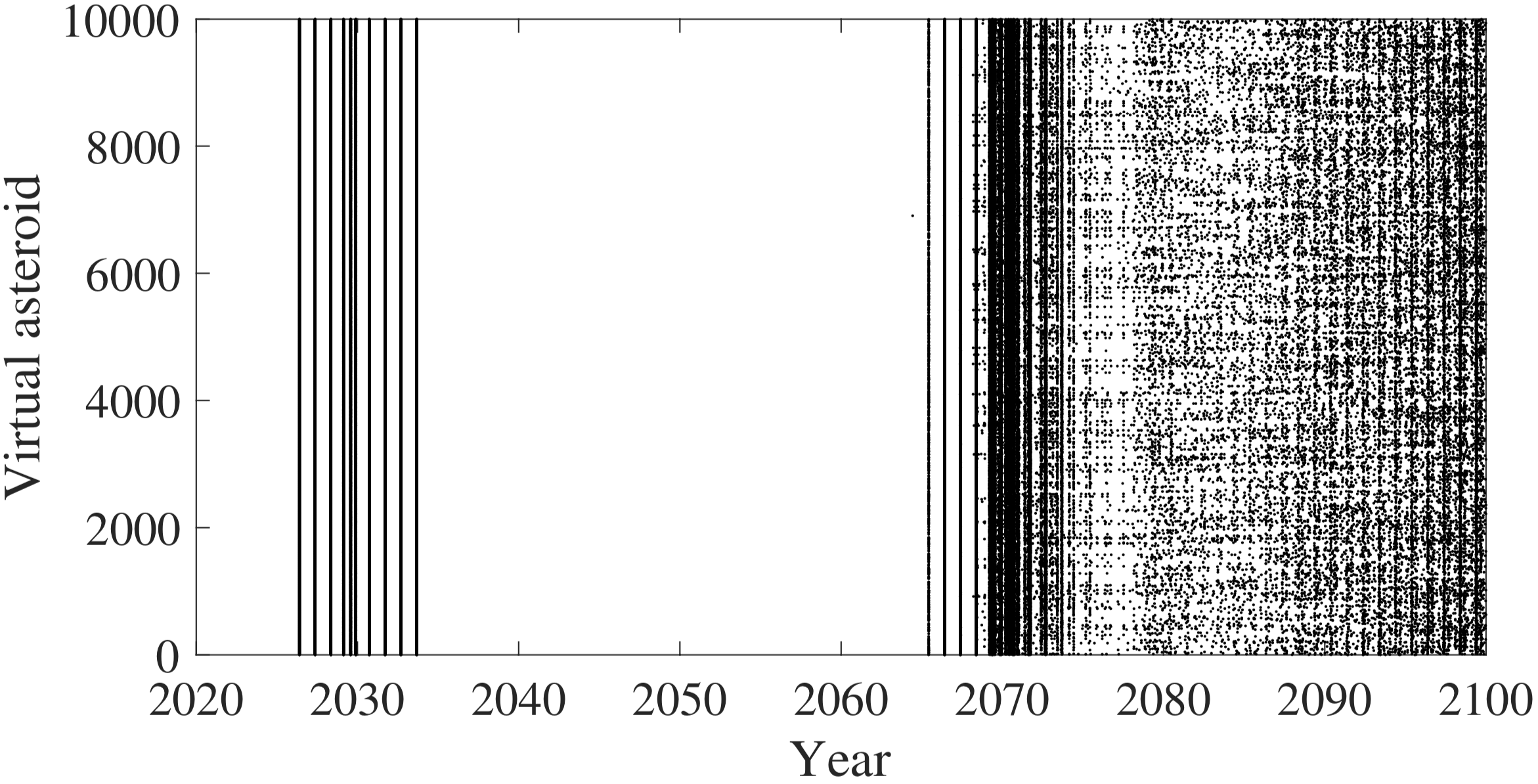}
    \caption{Distribution of close-approach dates (within 0.1~au) for asteroid 2000~SG$_{344}$ obtained with 10,000 virtual asteroids.}
    \label{Fig:ca_dates}
\end{figure}

As an example, Fig.~\ref{Fig:ca_dates} shows the distribution of close-approach dates obtained by propagating $N=10^4$ virtual asteroids sampled from the orbital uncertainty distribution of asteroid 2000~SG$_{344}$. Close approaches up to 2070 are clearly separated, and the sliding-window filter adequately clusters the different close approaches. However, the orbit of this asteroid is very close to Earth's \citep{chodas20012000} and, after a close encounter in 2071, there is a continuum of slow encounters ($v_\infty=1.4~\text{km/s}$) that makes it impossible for the sliding window to identify distinct nominal close approaches. In this case, close-approach clusters are split so that the same virtual asteroid can only appear once.

\subsubsection{Incorporating the impact pseudo-observation\label{Sec:find_VIs}}

%\subsubsection{Searching for virtual impactors}

Once all relevant close approaches have been identified, we search for impact trajectories within each cluster. Denoting $\myvec{b}_{s}(t^\ast)\in\mathbb{R}^2$ the coordinates on the scaled $b$-plane\footnote{\noindent See the Appendix and the references therein for a detailed definition of the scaled $b$-plane and the modified target plane.} \citep{chodas1999predicting} relative to Earth at a certain epoch $t^\ast$, the condition for a virtual asteroid to be an impactor, neglecting the flattening of the Earth, reduces to $||\myvec{b}_{s}(t^\ast)||<R_\earth$, where $R_\earth$ is the radius of the Earth extended to account for the atmosphere. When the orbit of the asteroid relative to Earth is not hyperbolic, as in the case of temporarily captured asteroids, the modified target plane is used in lieu of the $b$-plane \citep{yeomans1994predicting,milani1999asteroidII}. The impact condition can be treated as an observation by introducing the residuals
\begin{equation}\label{Eq:residual}
    \myvec{e}_b =  - \myvec{b}_{s}(t^\ast).
\end{equation}
The residuals $\myvec{e}_b$ are zero along an impact trajectory through the center of the Earth. To prevent numerical issues caused by the trajectory going through the center of attraction, a small fixed offset $\myvec{\varepsilon}$ is introduced ($||\myvec{\varepsilon}||\ll R_\earth$), transforming Eq.~\eqref{Eq:residual} into $\myvec{e}_b=\myvec{\varepsilon} - \myvec{b}_{s}(t^\ast)$.

The residuals $\myvec{e}_b$ are appended to the vector of observation residuals, $\myvec{e}$, incorporating the impact condition to the cost function in Eq.~\eqref{Eq:cost_od}. The uncertainty of the impact pseudo-observation is set to
\begin{equation}\label{Eq:sigma_b}
    \sigma_b = \frac{R_\earth}{\alpha},
\end{equation}
where $\alpha$ is a scaling factor that controls the size of the uncertainty ellipsoid on the scaled $b$-plane. Larger values of $\alpha$ overweight the impact condition over the observations and reduce the spread of the covariance on the scaled $b$-plane. By setting $\alpha=1$, for example, the a posteriori mapping of uncertainty from the extended filter results in a covariance ellipse on the scaled $b$-plane with a semimajor axis approximately equal to $R_\earth$ (1-sigma). The impact pseudo-observation typically constrains the orbit along the weak direction whereas the  data constrain the strong direction. 

The diagonal entries of the weight matrix corresponding to the residuals $\myvec{e}_b$ are set to $1/\sigma_b^2$. The corresponding entries of the design matrix are defined by the partial derivatives of the scaled $b$-plane coordinates with respect to $\myvec{x}$. The partial derivatives are computed from the numerically propagated variational equations \citep{milani2010theory} combined with the partial derivatives of the $b$-plane mapping \citep{farnocchia2019planetary}.

Incorporating the impact pseudo-observation replaces the one-dimensional search for impactors along the LOV performed by Sentry and CLOMON2. Instead of interpolating between two close approaches to find the impact trajectory, the filter, driven by the impact pseudo-observation, naturally marches through the uncertainty region toward the impactor.

\subsubsection{Selecting the initial guess}\label{Sec:initial_guess}
The convergence of the extended filter depends on the choice of the initial guess. The starting orbit must be close enough to the impact trajectory for the differential corrector to converge to it. The virtual asteroids that make a close-approach within 0.1~au (obtained in Section~\ref{Sec:ca_detection}) are the natural choice of initial guess.  Since the correction in Eq.~\eqref{Eq:diffcorr} follows the direction of the local gradient of the residuals, the initial guess must be dynamically connected to the impact trajectory following a continuous (differentiable) path across the basin of attraction. Under normal circumstances, the orbit computed by the filter follows the same dynamical path as the initial guess. 

%There may be one or more dynamical pathways leading to close approaches---and impacts, potentially---on a certain date. 

The brute-force approach to finding all possible VIs would be to run the extended filter starting from every virtual asteroid that makes a close approach. However, this is usually redundant and computationally inefficient because many virtual asteroids typically describe the same path and lead the filter to the same solution. Ideally, the filter should be run only once per dynamical path, starting from one representative virtual asteroid of each pathway. 

To identify each dynamical pathway, we group the samples making a close approach on a certain date (already identified using the sliding-window filter described in Sect.~\ref{Sec:ca_detection}) into clusters based on their dynamical evolution. Different clusters represent different dynamical paths leading to a close approach. Starting from the virtual asteroid with the earliest close-approach date in the list, we compare it to the rest of virtual asteroids and, for every pair, we evaluate a fitness function $\phi_{ij}$. If $\phi_{ij}\leq \phi_\text{max}$, the corresponding $(i,j)$ pair is clustered together. Then, we advance to the next virtual asteroid not yet belonging to any cluster and repeat the process. Maximizing the probability of finding a VI takes precedence over computational speed; therefore, the algorithm is not allowed to merge existing clusters because that could potentially connect distinct dynamical paths. This strategy results in a clustering algorithm with maximum complexity $O(N^2)$, which may become as low as $O(N)$ if all virtual asteroids are clustered at once into a single cluster.

The fitness function $\phi_{ij}$ is defined as follows. Let $\{\myvec{b}_{s}\}_k$ denote the scaled $b$-plane coordinates of the $k$-th virtual asteroid, $\{\myvec{x}\}_k$, at the time of close approach, $\{t_\text{ca}\}_k$, and let
\begin{equation}\label{Eq:partials_b_tca}
	\mymatrix{\Gamma}_k = \pder{\myvec{b}_{s}}{\myvec{x}}\bigg|_{\{\myvec{x}\}_{k}} \qquad \text{and} \qquad \myvec{\beta}_{k} = \pder{t_\text{ca}}{\myvec{x}}\bigg|_{\{\myvec{x}\}_{k}}
\end{equation}
denote the partial derivatives of the scaled $b$-plane coordinates and time of close approach with respect to $\myvec{x}$, respectively, evaluated along the trajectory of the $k$-th virtual asteroid. Given two virtual asteroids $\{\myvec{x}\}_{i}$ and $\{\myvec{x}\}_{j}$, the difference between their respective scaled $b$-plane coordinates and time of close approach can be estimated using the linear mapping defined in Eq.~\eqref{Eq:partials_b_tca}:
\begin{alignat}{2}
	& \{\Delta \myvec{b}_{s}\}_{ij}^\text{lin} && =  \mymatrix{\Gamma}_i \{\Delta\myvec{x}\}_{ij},\label{Eq:linear_deltab_ij}\\
	& \{\Delta t_\text{ca}\}_{ij}^\text{lin} && =  \myvec{\beta}_{i} \{\Delta\myvec{x}\}_{ij},
\end{alignat}
where $\{\Delta\myvec{x}\}_{ij} = \{\myvec{x}\}_{j} - \{\myvec{x}\}_{i}$. Similarly,
\begin{alignat}{2}
	& \{\Delta \myvec{b}_{s}\}_{ji}^\text{lin} && =  \mymatrix{\Gamma}_j \{\Delta\myvec{x}\}_{ji},\label{Eq:linear_deltab_ji}\\
	& \{\Delta t_\text{ca}\}_{ji}^\text{lin} && =  \myvec{\beta}_{j} \{\Delta\myvec{x}\}_{ji}.
\end{alignat}
The linear estimates can be compared with the actual differences obtained from the numerical propagation of each trajectory, 
\begin{alignat}{2}
 & \{\Delta\myvec{b}_{s}\}_{ij}^\text{non} && =\{\myvec{b}_{s}\}_j-\{\myvec{b}_{s}\}_i=-\{\Delta\myvec{b}_{s}\}_{ji}^\text{non},\\
 & \{\Delta t_\text{ca}\}_{ij}^\text{non} && =\{t_\text{ca}\}_j-\{t_\text{ca}\}_i = -\{\Delta t_\text{ca}\}_{ji}^\text{non}.
\end{alignat}

\begin{figure}
	\centering
	\includegraphics[width=.7\linewidth]{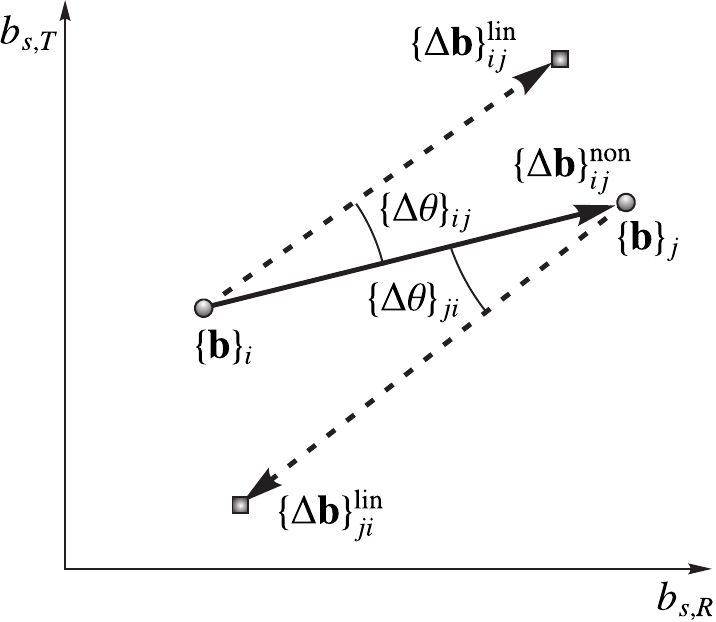}
	\caption{Schematic representation of the linear and nonlinear propagations to the scaled $b$-plane in polar coordinates, with $\{\Delta\theta\}_{ij} = |\{ \theta\}^{\text{lin}}_{ij} - \{ \theta\}^{\text{non}}_{ij}|$.\label{Fig:geometry_deltabplane}}
\end{figure}

Using polar coordinates on the scaled $b$-plane, $\{\Delta \myvec{b}_{s}\}_{ij}$ can be transformed into:
\begin{alignat}{2}
	& \{ \rho\}_{ij} && = ||\{\Delta \myvec{b}_{s}\}_{ij}||, \label{Eq:polar_rho}\\
	& \{ \theta\}_{ij} && = \arctan\frac{\{\Delta b_{s,T}\}_{ij}}{\{\Delta b_{s,R}\}_{ij}} \label{Eq:polar_theta}
\end{alignat}
based on Eq.~\eqref{Eq:components_b_scaled}. The error functions
\begin{alignat}{2}
	& \{E_\rho\}_{ij} &&=\frac{1}{\{ \rho\}^\text{non}_{ij}}|\{ \rho\}^\text{lin}_{ij} - \{ \rho\}^\text{non}_{ij}|, \label{Eq:diff_rho}\\
	& \{E_\theta\}_{ij} &&= \frac{1}{\theta_\text{max}}|\{ \theta\}^{\text{lin}}_{ij} - \{ \theta\}^{\text{non}}_{ij}|, \label{Eq:diff_theta}\\
	& \{E_{t_\text{ca}}\}_{ij} &&=\frac{1}{|\{ \Delta t_\text{ca}\}^\text{non}_{ij}|} |\{ \Delta t_\text{ca}\}^\text{lin}_{ij} - \{\Delta  t_\text{ca}\}^\text{non}_{ij}|,\label{Eq:diff_tca}
\end{alignat}
quantify how much the linear map deviates from the numerical propagation to the $b$-plane. The difference in Eq.~\eqref{Eq:diff_theta} is defined as the shortest angular distance and the maximum angular separation is set to $\theta_\text{max}=20^\circ$. Lower bounds on $\{ \rho\}^\text{non}_{ij}$ and $\{ \Delta t_\text{ca}\}^\text{non}_{ij}$ are set to prevent singularities. Equations~(\ref{Eq:diff_rho}--\ref{Eq:diff_tca}) lead to the fitness function
\begin{equation}\label{Eq:cost_J}
	\phi_{ij} = \max(\{E_\theta\}_{ij}, \{E_\rho\}_{ij}, \{E_{t_\text{ca}}\}_{ij}, \{E_\theta\}_{ji}, \{E_\rho\}_{ji}, \{E_{t_\text{ca}}\}_{ji})
\end{equation}
and we set $\phi_\text{max}=1$. Figure~\ref{Fig:geometry_deltabplane} sketches the geometry of the comparison of a pair of virtual asteroids on the $b$-plane.

The maximum value of $\phi_{ij}$ observed when producing a cluster is an indicator of the compactness of that cluster. Small values of $\phi_{ij}$ indicate that the cluster is very compact, meaning that the cluster is well defined, clearly distinct from others, and that the samples evolve in the linear regime. As a result, the filter converges to the same orbit starting from any  virtual asteroid in the cluster. Conversely, large values of $\phi_{ij}$ indicate loose clusters, which may not be well defined. In this case, the filter tests a large number of initial guesses per close-approach, increasing the probability of finding a VI.

\citet{roa2019multilayer} presented an alternative approach to clustering the virtual asteroids in MC simulations by looking for patterns in parameter space directly. However, this approach requires defining a metric to rank each pair of samples together with a cutoff threshold to determine when to cluster them or, equivalently, defining the number of clusters before hand. Defining these parameters robustly while still obtaining good computational performance is quite challenging. For this reason, we opted for the new clustering algorithm based on the fitness function in Eq.~\eqref{Eq:cost_J}. The main advantage of the proposed clustering technique is that the value of $\phi_\text{max}$ has a physical interpretation based on the close-approach parameters. When the dynamics are linear and the clusters are compact, the algorithm reduces the number of initial guesses to be tested to essentially one sample per close approach. This is usually the case for early close approaches. As time progresses and the orbit undergoes close planetary encounters, gravitational scattering can produce multiple dynamical paths to branch out, resulting in the algorithm finding multiple clusters for a given close-approach date. Finding multiple clusters per close approach causes the filter to more thoroughly explore the space of parameters, minimizing the chances of missing a VI in pathological cases when the dynamics are nonlinear or the clusters are not clearly differentiated.

%in pathological cases when the dynamics are nonlinear or the clusters are not clearly differentiated, the algorithm finds multiple clusters per close approach, causing the filter to more thoroughly explore the space of parameters, which minimizes the chances of missing a VI.

%\new{The clustering technique presented in this section groups samples that make a specific close approach following the same dynamical path. For early close-approach dates, all samples are usually grouped in a single cluster. As time progresses and the orbit undergoes close planetary encounters, gravitational scattering can produce multiple dynamical paths to branch out, resulting in the algorithm finding multiple clusters for a given close-approach date.}

\subsubsection{Characterizing the VI}\label{Sec:char_VI}

%The filter extended with the impact pseudo-observation returns the nominal value of the parameters defining an impact trajectory that is consistent with the observations, $\bar{\myvec{x}}_{i}$, using Eq.~\eqref{Eq:diffcorr}.

The extended filter solves Eq.~\eqref{Eq:diffcorr} to find the orbit that minimizes the observation residuals while impacting Earth on a given date. The solution is the nominal orbit of the VI, $\bar{\myvec{x}}_{i}$. The covariance of the fit, $\mymatrix{\Sigma}_{i}$, is given by Eq.~\eqref{Eq:covariance_filter} incorporating the impact pseudo-observation and approximates the shape of the region in parameter space that leads to impacts. The $\alpha$-sigma covariance is a representation of the VI $F_i$. If the filter fails to converge, it generally means that the observational data do not admit impact trajectories within the confidence level defined by $\alpha$ in Eq.~\eqref{Eq:sigma_b}. The filter may converge to a trajectory that comes closer to Earth than the initial guess but that is still not an impact trajectory, i.e., the data admit closer approaches but not impacting solutions. If the nominal VI orbit $\bar{\myvec{x}}_i$ falls outside ${D}_7$, we discard the VI because of not being statistically significant.

The spread of the covariance ellipsoid of the VI is substantially controlled by the parameter $\alpha$ that weights the impact pseudo-observation, as shown in Eq.~\eqref{Eq:sigma_b}. Experience shows that an initial run with a relatively large value of $\alpha$ (we set $\alpha=10$ for Sentry-II) pulls the initial guess strongly toward an impact trajectory, which is desirable when trying to find VIs. When modeling the VI with the p.d.f.\ obtained with such a large value of $\alpha$, most virtual asteroids sampled from $p_i(\myvec{x})$ are impactors and $\Sigma_i$ might be underestimating the actual extent of the VI. For a better description of the VI, once the impact trajectory has been found, the filter is run a second time starting from the previously converged solution using a looser value of $\alpha$ to determine the actual extent of the VI covariance but without correcting the orbit solution. For Sentry-II, we choose $\alpha$ adaptively to make the uncertainty of the impact pseudo-observation equal to half of the chord of the Earth's cross section given the nominal impact location. Scaling $\alpha$ adaptively compensates for VIs with nominal impact locations far from the geocenter.

\begin{figure}[t]
    \centering
\gridline{\fig{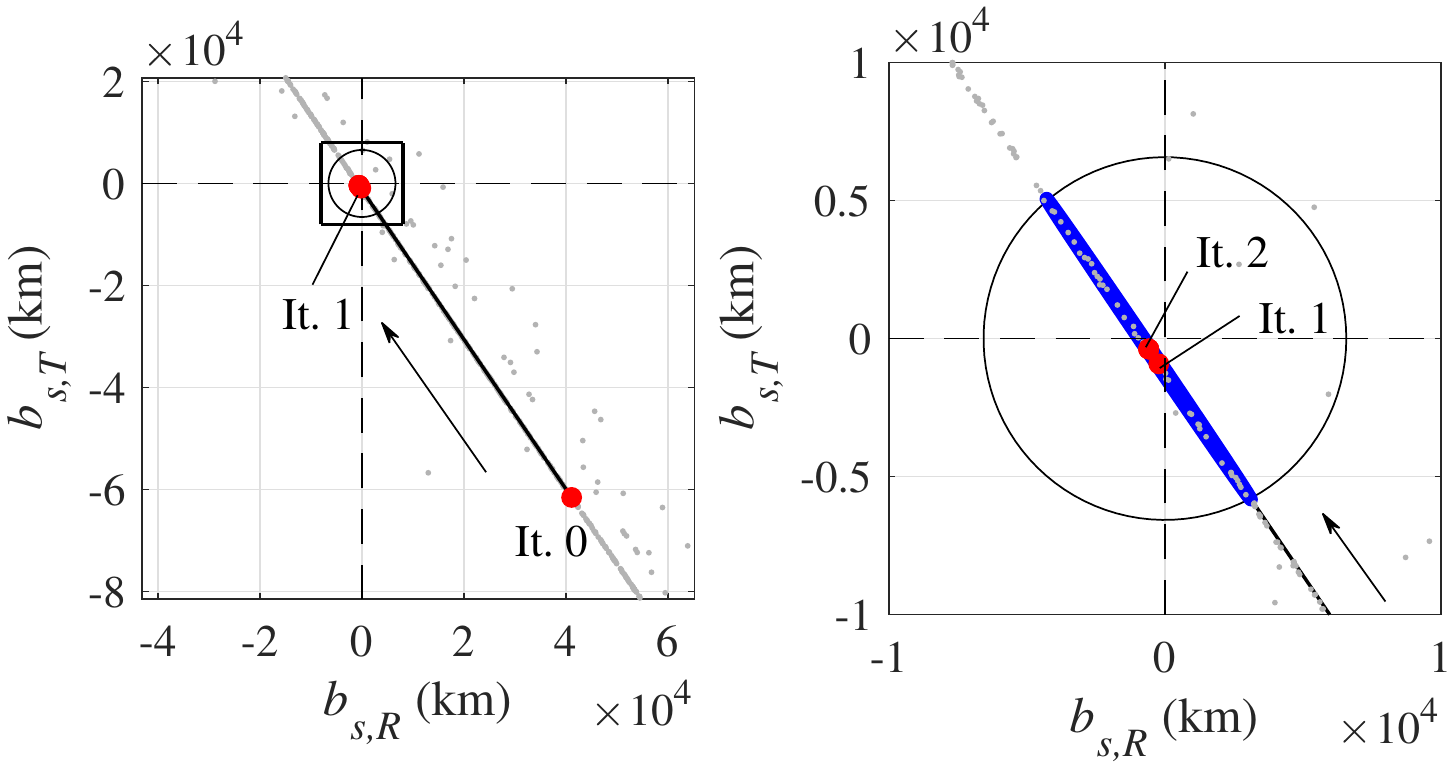}{\linewidth}{(a) Impact on 2080-10-08.25, $P(F_i)=4.8\times10^{-5}$.}}
\gridline{\fig{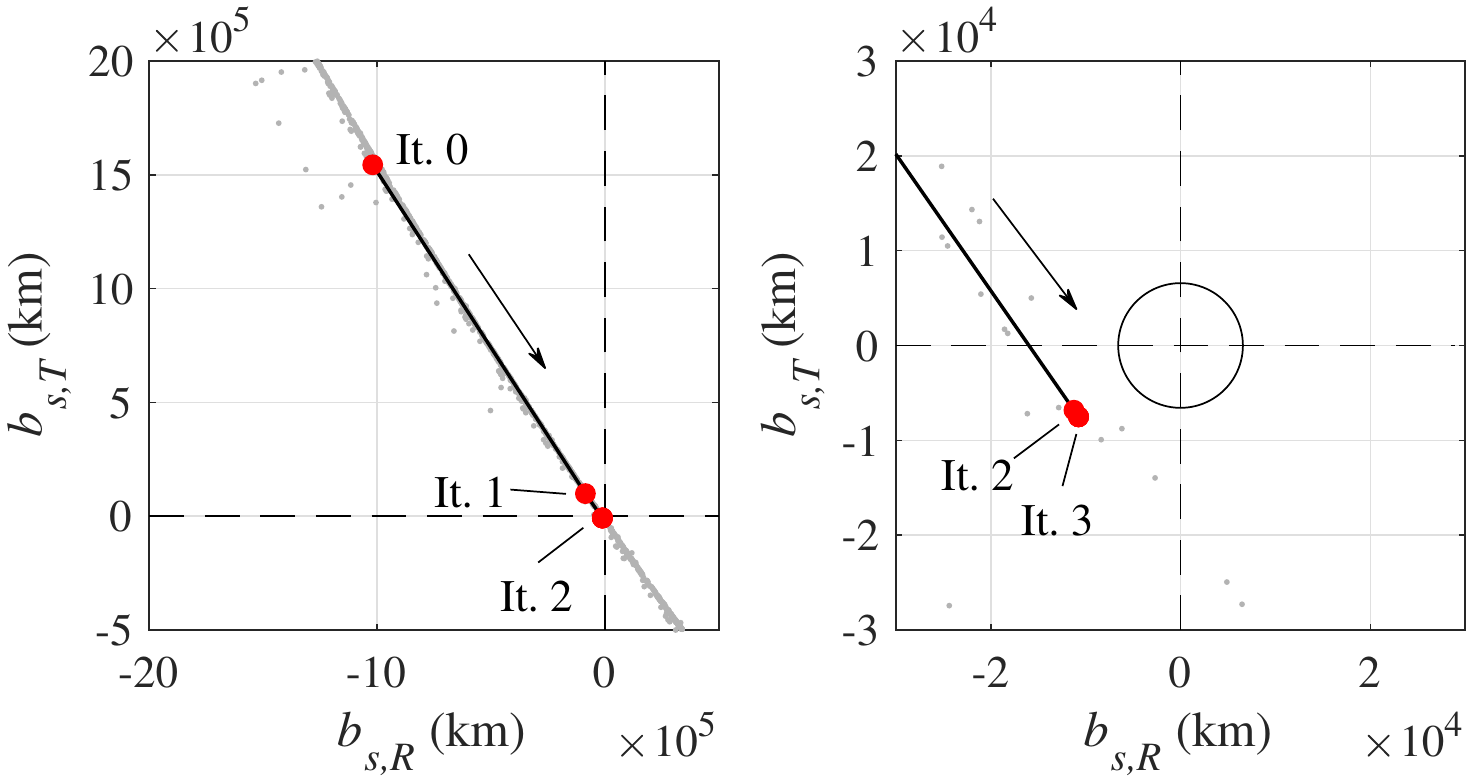}{\linewidth}{(b) Near miss on 2073-10-08.20.}}
    \caption{Convergence sequence on the scaled $b$-plane for two different close approaches of asteroid 2000~SB$_{45}$. The right panel is a zoomed-in view centered at the Earth. Gray dots represent MC samples drawn from $q(\myvec{x})$ that outline the weak direction. The blue ellipse on the top-right panel is the 1-sigma covariance ellipse. The black circumference is the Earth cross section.\label{Fig:converge}}
\end{figure}

Figure~\ref{Fig:converge} illustrates how the extended filter including the impact pseudo-observation works in practice. It shows the evolution of the point of closest approach on the scaled $b$-plane for the 2080 and 2073 close approaches of asteroid 2000~SB$_{45}$. As a reference, the gray dots represent MC samples drawn from $q(\myvec{x})$ and propagated numerically to the scaled $b$-plane. Note that not all the samples are dynamically linked. The close approach in 2080 (Fig.~\ref{Fig:converge}a) admits a VI with impact probability $4.8\times10^{-5}$. Starting from an initial guess (``It.~0'') that makes a close approach at approximately 75,000~km from the geocenter, the filter advances along the weak direction (outlined by the stream of gray dots obtained via MC sampling) and the convergence is substantially complete after two iterations. The zoomed-in view on the right panel shows the projection on the scaled $b$-plane of the 1-sigma covariance ellipsoid of the VI obtained by setting $\alpha\approx1$ in Eq.~\eqref{Eq:sigma_b}. The semimajor axis of the 1-sigma ellipse spans approximately one Earth radius. The close approach in 2073 (Fig.~\ref{Fig:converge}b) results in a near miss. The stream of MC samples shows how the uncertainty region mapped to the scaled $b$-plane is close to Earth but the observational data constrains the strong direction in a way that impacts are not possible. The filter follows the weak direction and converges to a close approach at approximately 10,000~km from the geocenter. %This is an additional feature of the proposed technique because it can find the solution coming closest to Earth even when an impact is not possible.

% 2080-10-08 2000 SB45

\begin{figure}[t]
    \centering
    \includegraphics[width=\linewidth]{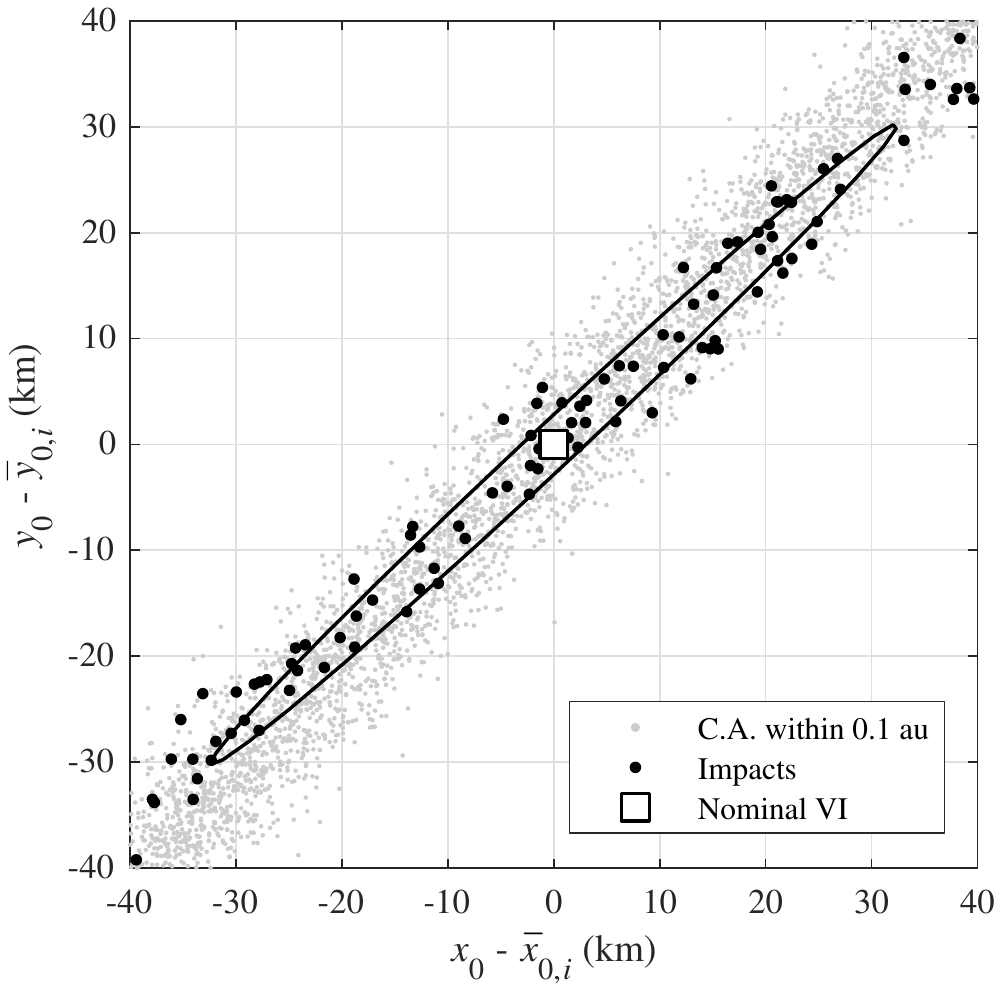}
    \caption{Projection in initial-conditions space of the 1-sigma covariance ellipsoid of the 2022-May-06 VI of asteroid 2009~JF$_1$. The square marker represents the nominal orbit of the VI, gray dots are the subset of virtual asteroids that make a close approach within 0.1~au on this date from a total of $10^6$ MC samples drawn from $q(\myvec{x})$, and the black dots represent impacting solutions.}
    \label{Fig:vi_2009jf1}
\end{figure}

Figure~\ref{Fig:vi_2009jf1} shows the projections on the $x_0$--$y_0$ plane of the orbital uncertainty distribution of the VI from a different example, the 2022-May-06 VI of asteroid 2009~JF$_1$.  The higher impact probability of this VI ($2.6\times10^{-4}$) makes it easier to visualize. The ellipse is the projection of the 1-sigma covariance ellipsoid of the VI computed by the filter, $\mymatrix{\Sigma}_{i}$, which outlines the region in parameter space where dynamically connected impacting orbits may be found. Impacting solutions are represented by black dots. The semimajor axis of the projected VI covariance is approximately 5\% of that of the initial orbital uncertainty distribution. The covariance obtained by the filter approximates the region in parameter space where impacts occur.  %Sampling a set of virtual asteroids directly from $p_i(\myvec{x})$ and propagating them to the $b$-plane results in a 95\% impact ratio.

\subsection{Estimating the impact probability}\label{Sec:ip_is}
Once the nominal orbit $\bar{\myvec{x}}_{i}$ and the covariance matrix ${\mymatrix{\Sigma}}_{i}$ of a VI have been obtained using the extended filter, the impact probability can be estimated from Eq.~\eqref{Eq:ip_definition}. This integral expression cannot be solved analytically. Instead, the impact probability is approximated numerically. For relatively large impact probabilities, direct MC sampling starting from the initial distribution of orbital uncertainty $q(\myvec{x})$ provides a robust probability estimate. However, MC sampling proves inefficient when the impact probability is small because at least a few times $1/P(F_i)$ samples must be propagated to the time of close approach.

The inefficiency of MC sampling for estimating small probabilities relates to the fact that most of the samples drawn from the initial probability distribution $q(\myvec{x})$ fall outside the region of interest $F_i$. Therefore, a potentially prohibitively large number of samples must be drawn to model this region with adequate resolution. To overcome this limitation, importance sampling introduces a proposal p.d.f.\ to sample from, $p_i(\myvec{x})$, which substantially increases the ratio of samples that fall inside $F_i$ \citep[see, for example,][]{ripley2009stochastic}. That is, sampling from $p_i(\myvec{x})$ increases the ratio of impacts compared to sampling from $q(\myvec{x})$. Denoting $N_\text{ip}$ the number of samples to be drawn from $p_i(\myvec{x})$, the probability estimate can be transformed into
\begin{alignat}{1}
    P(F_i) &= \int_{F_i} p_i(\myvec{x})\frac{q(\myvec{x})}{p_i(\myvec{x})}\,\diff\myvec{x} 
    \approx \frac{1}{N_{\text{ip}}}\sum_{k=1}^{N_{\text{ip}}} I_{F_i}(\{\myvec{x}\}_k)w_{i,k},\label{Eq:ip_is}
\end{alignat}
which is written in terms of the likelihood ratio
\begin{equation}
	w_{i,k} = \frac{q(\{\myvec{x}\}_k)}{p_i(\{\myvec{x}\}_k)}.
\end{equation}
The $\{\myvec{x}\}_{k=1,\ldots,N_\text{ip}}$ virtual asteroids are sampled from the proposal distribution $p_i(\myvec{x})$. The standard deviation associated with the probability estimate follows from the variance of the estimate \citep{owen2000safe,rubino2009rare}:
\begin{equation}
    %\sigma_{P_i}^2 =   \frac{1}{N_\text{ip}^2} \sum_{k=1}^{N_\text{ip}} [I_{F_i}(\{\myvec{x}\}_k)w_{i,k} - P(F_i)]^2 .
    \sigma_{P_i}^2 =   \frac{1}{N_\text{ip}} \Bigg\{ \frac{1}{N_\text{ip}}\sum_{k=1}^{N_\text{ip}} \big[I_{F_i}(\{\myvec{x}\}_k)w_{i,k}^2\big] - \big[P(F_i)\big]^2 \Bigg\}.
\end{equation}

In general, the most challenging step when estimating the impact probability using importance sampling is finding an adequate proposal distribution $p_i(\myvec{x})$ to model each VI. Incorporating the impact pseudo-observation into the filter solves this problem entirely as the covariance of the impacting solution naturally models the VI, outlining the region in parameter space that produces impacts along the same dynamical path. We define the proposal $p_i(\myvec{x})$ as
\begin{equation}\label{Eq:pdf_VI}
    p_i(\myvec{x})=\mathcal{N}(\bar{\myvec{x}}_{i},\mymatrix{\Sigma}_{i})
\end{equation}
in terms of the nominal orbit $\bar{\myvec{x}}_i$ and covariance $\mymatrix{\Sigma}_i$ of the VI.

To evaluate the indicator function $I_{F_i}(\{\myvec{x}\}_k)$, the corresponding orbit must be propagated to the time of close approach to check if it impacts the Earth or not. Assuming that the dynamics are linear within $F_i$, which is generally a small subset of ${D}_\sigma$, the numerical integration of each trajectory can be replaced with the far more efficient linear mapping of the parameters onto the scaled $b$-plane,
\begin{equation}\label{Eq:linear_bmap}
    \{\myvec{b}_{s}\}_k = \bar{\myvec{b}}_{s,i} + \bar{\mymatrix{\Gamma}}_i\{\Delta \myvec{x}\}_k,\qquad \bar{\mymatrix{\Gamma}}_i = \pder{\myvec{b}_{s}}{\myvec{x}}\bigg|_{\bar{\myvec{x}}_{i}},
\end{equation}
where $\bar{\myvec{b}}_{s,i}$ denotes the scaled $b$-plane coordinates of the nominal trajectory of the VI, $\{\Delta \myvec{x}\}_k = \{ \myvec{x}\}_k - \bar{\myvec{x}}_{i}$, and $\mymatrix{\Gamma}_i$ contains the partial derivatives of the scaled $b$-plane coordinates with respect to the parameters along the nominal trajectory of the VI. For Sentry-II, we choose $N_\text{ip}=10^5$.

Section~\ref{Sec:nonlinear} presents a method for testing the hypothesis of linearity within $F_i$. When the linear approximation no longer applies based on the test described in Section~\ref{Sec:nonlinear}, Sentry-II evaluates Eq.~\eqref{Eq:ip_is} using MC sampling by numerically propagating the orbits of $N_\text{ip}=200$ virtual asteroids sampled from $p_i(\myvec{x})$. % Given the computational efficiency of the linear mapping, the probability estimate in Eq.~\eqref{Eq:ip_is} can be evaluated using a large number of samples, e.g. $N_\text{ip}=10^4$, with minimal overhead.

%\textcolor{red}{Add a figure showing impacting samples during the IS step in ics and b-plane. Use one example with two VIs on the same date}

Equation~\eqref{Eq:linear_bmap} must produce a balanced distribution of impact and non-impact trajectories for the estimator in Eq.~\eqref{Eq:ip_is} to be unbiased. If all samples result in impacts, the VI covariance can be underestimating the actual extent of the VI. On the other hand, if too few samples are impactors, the VI will not be characterized properly due to a poor sampling resolution. The impact ratio can be controlled with the selection of $\alpha$. Impact ratios between 40--80\% work well in practice, which can generally be achieved by setting $\sigma_b$ equal to half of the chord of the Earth's cross section when recomputing the final VI covariance. %Nonetheless, the impact ratio may be lower when the impact location on the scaled $b$-plane is not near the geocenter and the chord spanned by the covariance ellipse is shorter than Earth's diameter. If the impact ratio is too low, the a posteriori VI covariance is scaled down to increase the number of impactors.

\section{Practical Aspects of the Implementation}\label{Sec:discussion}

\subsection{Nonlinearity test\label{Sec:nonlinear}}
%When estimating the impact probability using importance sampling, the initial conditions sampled from each $p_i(\myvec{x})$ are propagated to the $b$-plane using the linear mapping in Eq.~\eqref{Eq:linear_bmap}. If the dynamics of the VI is nonlinear in this particular time span (due to, for example, a very close planetary encounter resulting in the gravitational scattering of nearby trajectories) the probability estimate may not be accurate. This section devises a nonlinearity test to be run on every converged VI to assess the validity of the approximation.

The extended filter implicitly assumes that the dynamics within the VI is linear. Although this assumption holds in the vast majority of cases, the present section devises a nonlinearity test to be run on every converged VI to assess its validity. The nonlinearity test is only run on the impacting solutions because we are interested in assessing how well the linear assumption holds inside the Earth cross section.

A small number of virtual asteroids is sampled randomly from the uncertainty distribution of each VI (Sentry-II samples 20~virtual asteroids) and they are propagated numerically to the scaled $b$-plane to obtain a fully nonlinear solution. To increase the impact ratio, the VI covariance is scaled down by a factor of two when sampling the virtual asteroids. Sampling the virtual asteroids randomly as opposed to using a deterministic set of sigma points simplifies the test because no scaling parameters are required. Let $\{\Delta\myvec{b}_{s}\}_k^{\text{non}}$ denote the separation on the scaled $b$-plane between the $k$-th virtual asteroid and the nominal $b$-plane coordinates of the VI. Each virtual asteroid is mapped to the scaled $b$-plane again only this time using the linear mapping in Eq.~\eqref{Eq:linear_bmap}, producing $\{\Delta\myvec{b}_{s}\}_k^{\text{lin}}$. The linear and nonlinear deviations are converted to polar coordinates using Eqs.~(\ref{Eq:polar_rho}--\ref{Eq:polar_theta}), resulting in $\{\rho\}^\text{non}_k$, $\{\theta\}^\text{non}_k$, $\{\rho\}^\text{lin}_k$, $\{\theta\}^\text{lin}_k$. Using the error functions in Eqs.~(\ref{Eq:diff_rho}--\ref{Eq:diff_theta}), we define the nonlinearity index:
\begin{equation}\label{Eq:linearity}
	\ell = \max_k(\{E_\rho\}_k, \{E_\theta\}_k).
\end{equation}

The criterion in Eq.~\eqref{Eq:linearity} is sensitive to roundoff errors when the asteroid undergoes strong planetary encounters. Small numerical errors that get amplified over time can result in errors of hundreds of kilometers when mapped onto the $b$-plane at a future date. To improve the robustness of the nonlinearity test in these situations, we implement a second test based on the partial derivatives of the scaled $b$-plane coordinates with respect to the initial conditions along each of the propagated samples. Denoting $\varGamma_{ij,m}$ the $(i,j)$ component of the matrix of partial derivatives $\mymatrix{\Gamma}_m$ in Eq.~\eqref{Eq:partials_b_tca}, we evaluate an additional nonlinearity index:
\begin{equation}\label{Eq:linearity_partials}
	\ell = \frac{1}{\ell_\text{max}}\min(\ell_{mn},\ell_{nm}),
\end{equation}
defined by the pair-wise ratio
\begin{equation}
	\ell_{mn} = \max_{ij} \frac{|\varGamma_{ij,m} - \varGamma_{ij,n}|}{|\varGamma_{ij,n}|},
\end{equation}
to compare the partial derivatives along two propagated samples $\{\myvec{x}\}_m$ and $\{\myvec{x}\}_n$. The scaling factor in Eq.~\eqref{Eq:linearity_partials} is set to $\ell_\text{max}=0.5$. For a VI to be flagged as nonlinear, both Eqs.~\eqref{Eq:linearity} and~\eqref{Eq:linearity_partials} must satisfy $\ell>1$.

In extremely nonlinear cases, only a small fraction of the virtual asteroids sampled from $F_i$ may impact Earth and some may not even come within 0.1~au of the Earth. In that case, the VI is immediately flagged as nonlinear.

%If the linearity index
%\begin{equation}\label{Eq:linearity}
%    \ell = \max_k\frac{||\{\Delta\myvec{b}_{s}\}_k^\text{lin} - \{\Delta\myvec{b}_{s}\}_k^\text{non}||}{||\{\Delta\myvec{b}_{s}\}_k^\text{non}||},
%\end{equation}
%which quantifies the error of the linear mapping, is greater than unity for a particular VI, that VI is flagged as nonlinear. 

%In such cases, the probability estimate obtained with Eqs.~\eqref{Eq:ip_is} and~\eqref{Eq:linear_bmap} may not be accurate and should be handled with care. Furthermore, nonlinear VIs may not be well characterized by the covariance in parameter space $\mymatrix{\Sigma}_{i}$ obtained by the extended filter. The reason is that the covariance is aligned with the tangent subspace spanned by the local gradient and it may fail to capture the curvature of a nonlinear region in parameter space. 

%These issues are aggravated when the orbit transitions to chaotic regime.

Strongly nonlinear dynamics can have further implications. First, the extended filter may struggle to converge to an impact solution from a specific seed. The differential corrector is ultimately a gradient-based method and the corrections must be small enough to remain inside the basin of attraction, which may be very small and have a complicated shape. In practice, this problem can be solved by limiting the magnitude of the correction in Eq.~\eqref{Eq:diffcorr} to improve robustness at the cost of penalizing the rate of convergence. Section~\ref{Sec:limit} explains the details of how the correction is bounded.

Second, once an impact trajectory is found, its covariance may not represent the VI accurately because it may have an irregular, nonlinear shape in parameter space.  The covariance is aligned with the tangent subspace spanned by the local gradient and it fails to capture the curvature of a nonlinear region in parameter space. In such situations, the proposed impact-monitoring system is expected to identify the VIs, but characterizing them could require special sampling techniques, like, for example, Markov-chain Monte Carlo \citep{cowles1996markov,muinonen2012asteroid}. 

Third, evaluating the indicator function $I_{F_i}$ over the virtual asteroids $\{\myvec{x}\}_k$ using the linear method in Eq.~\eqref{Eq:linear_bmap} may not be accurate and introduce errors in the probability estimate. This particular problem is solved by replacing the linear mapping in Eq.~\eqref{Eq:linear_bmap} with the numerical propagation of a set of virtual asteroids sampled from $p_i(\myvec{x})$ when the VI fails the linearity test.

Fourth, the numerical propagation of the trajectory may not be an accurate representation of the true motion of the asteroid. Numerical errors accumulate over time and strong nonlinearities can cause the integrator to diverge exponentially fast from the trajectory that it was supposed to propagate (see \citet{quinlan1992reliability} for a dedicated study of the reliability of generic $N$-body integrations). Regularization techniques may partially alleviate this issue and extend the validity of the integration \citep{stiefel1971linear,bond1996modern,roa2017regularization}. This problem is common to any impact-monitoring technique irrespective of how the VIs are characterized.

Even in the presence of strong nonlinearities, the nominal VI orbit and the impact date found by the filter still represent real impact conditions (limited by the fidelity of the orbit propagator and the force model within the requested time span). Therefore, the proposed method successfully identifies the nominal VIs but their uncertainty and the impact probability estimate may need to be refined in pathological cases.

%The proposed technique may underestimate the impact probability in these situations because the covariance $\mymatrix{\Sigma}_{i}$ captures only a fraction of the VI in parameter space. %Section~\ref{Sec:merging} discusses how combining several covariance ellipsoids may improve the probability estimate in nonlinear cases.

%It should be noted that, 

\subsection{Limiting the magnitude of the correction\label{Sec:limit}}
The magnitude of the correction in Eq.~\eqref{Eq:diffcorr} is limited to improve the robustness of the filter when dealing with nonlinearities. We limit the correction by imposing an a priori constraint on the uncertainty of the perihelion distance, eccentricity, and time of perihelion passage. The a priori uncertainty is set to 1\% of the formal, marginal uncertainty of each element. If there are other estimated parameters in addition to the state vector, their a priori uncertainties are also limited to 1\% of their formal uncertainties. The a priori constraints are reset at each iteration, helping the extended filter advance guided by the impact pseudo-observation.

Limiting the magnitude of the correction may result in the filter getting trapped in a local minimum and not finding the global optimum. To prevent this issue, once the filter has converged to an impact or a near miss, the filter is run again without any a priori constraint starting from the converged solution. If the unconstrained solution fails to converge, the filter is run again sequentially with a priori constraints set to 50\% and then 10\% of the formal uncertainties until it converges to an impact.

After identifying the impact solution with the constrained or the unconstrained filter passes, the filter is run one last time without updating the orbit to produce the unconstrained covariance matrix of the VI setting $\alpha$ to make $\sigma_b$ equal to half of the chord of the Earth's cross section, as indicated in Section~\ref{Sec:char_VI}.

\subsection{Merging virtual impactors}\label{Sec:merging}
It is possible for the extended filter to converge to the same VI starting from different initial guesses if they all lie in the same basin of attraction. In fact, when the dynamics are linear and there is only one VI per close-approach date, every initial guess will lead to this same VI. For this reason, the algorithm must be able to detect repeated VIs and to connect regions in parameter space that follow the same dynamical path.

Once all the close approaches have been processed and a preliminary list of VIs is available, each VI is compared pairwise with the rest of VIs in the list to determine if they are actually the same one and if they should be merged. Merging two VIs reduces to retaining the most representative one and discarding the other. The most representative VI is the one that was found with the loosest a priori constraints when running the extended filter. If both VIs were found with the same a priori constraint, the most representative one would be the one that comes closest to the geocenter.

%We propose two algorithms to decide if a pair of VIs needs to be merged: one for the case when both VIs are linear, and another one for the case when at least one of the VIs is nonlinear according to Section~\ref{Sec:nonlinear}.

There are three steps in the merging process. First, to ensure that the two VIs impact the Earth on the same date, their nominal impact dates, $t^\ast_i$ and $t_j^\ast$, must satisfy the condition
    \begin{equation}
        \max_{k=i,j}(|t^\ast_i - t^\ast_j| / \sigma_{t^\ast,k}) < m_t,
    \end{equation}
    where $\sigma_{t^\ast,k}$ is the uncertainty in the close-approach date of the $k$-th VI. The threshold $m_t$ controls the confidence level. For Sentry-II, we set $m_t=2$. If this criterion is not met, the VIs are not merged.
    
Second, we compare the VIs in parameter space. Defining their Mahalanobis distance in parameter space as
\begin{equation}
    d_{ij} = \sqrt{(\bar{\myvec{x}}_{i} - \bar{\myvec{x}}_{j})^\top\mymatrix{\Sigma}_{i}^{-1}(\bar{\myvec{x}}_{i} - \bar{\myvec{x}}_{j})},
\end{equation}
we impose the condition
\begin{equation}\label{Eq:condition_mahalanobis}
	\min(d_{ij},d_{ji})<m
\end{equation}
in terms of a constant parameter that we set to $m=2$. If a pair of VIs satisfy the condition in Eq.~\eqref{Eq:condition_mahalanobis}, they are merged. This criterion applies only to the case when both VIs are flagged as linear.
    
Nonlinear VIs may not be well characterized by their covariance in parameter space so their mutual Mahalanobis distance is not always an adequate metric. To connect dynamical paths for those cases where Eq.~\eqref{Eq:condition_mahalanobis} does not hold, we introduce a third criterion based on the impact geometry. As part of the nonlinearity test described in Section~\ref{Sec:nonlinear}, we already propagated a small set of virtual asteroids sampled from the orbital uncertainty distribution of every VI to the scaled $b$-plane. If $\ell<1$ for any pair of virtual asteroids each sampled from a different VI, where $\ell$ is defined in Eq.~\eqref{Eq:linearity_partials}, the parent VIs are merged. This criterion assumes that two virtual asteroids with the same partial derivatives of their scaled $b$-plane coordinates with respect to $\myvec{x}$ follow the same dynamical path. In other words, small deviations in parameter space produce the same effect on both trajectories.

\begin{figure}[h]
	\centering
	\includegraphics[width=\linewidth]{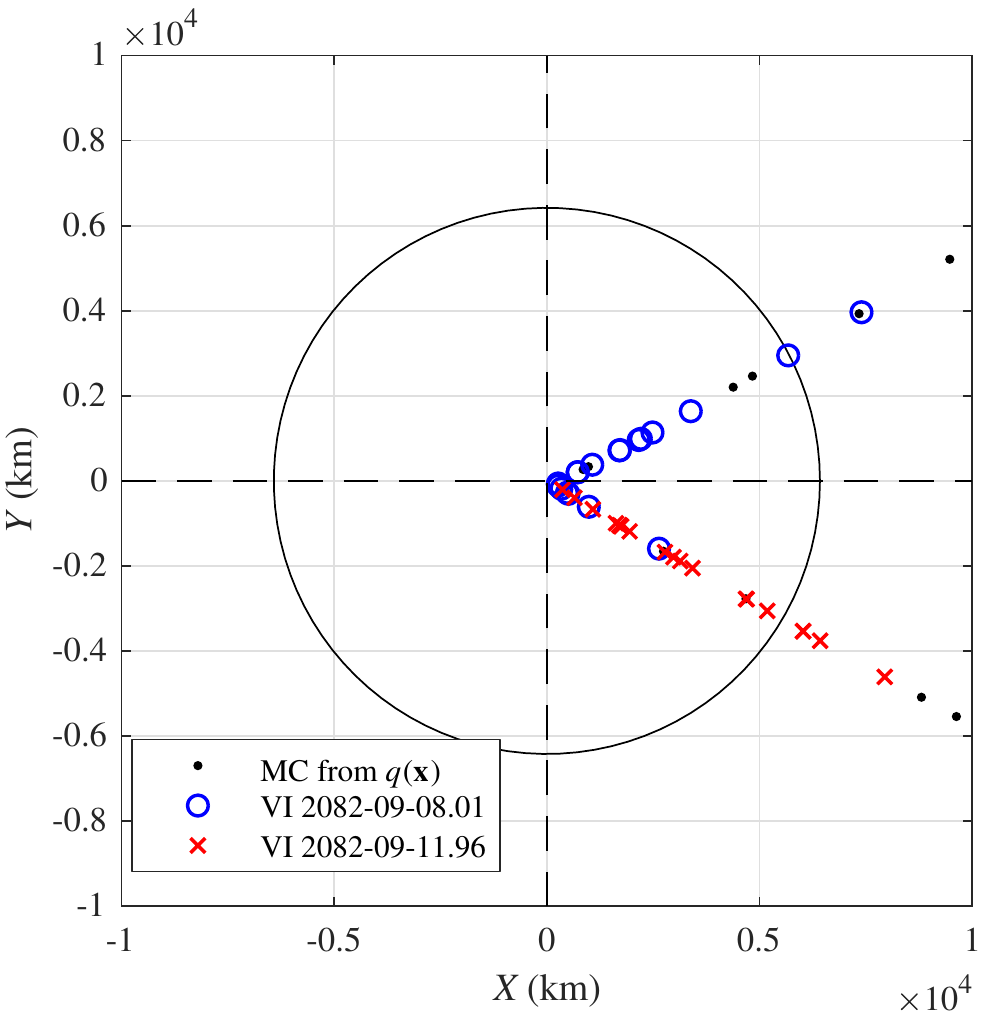}
	\caption{Coordinates on the modified target plane of the set of virtual asteroids used to merge the VIs detected during the September 2082 encounter of asteroid 2020~CD$_3$. Circular markers represent the virtual asteroids sampled from the uncertainty distribution of the VI found by the extended filter when starting from an initial guess on the upper branch, while crosses correspond to the VI found from a seed on the lower branch.\label{Fig:fold_2020cd3}}
\end{figure}

Figure~\ref{Fig:fold_2020cd3} represents the modified target plane of the September 2082 encounter of asteroid 2020~CD$_3$ \citep{naidu2021precovery}. The modified target plane is used in this case because the orbit is not hyperbolic with respect to Earth during the close encounter. The orbit solution includes estimated parameters to model nongravitational forces, which the proposed method treats systematically. An MC simulation using $10^6$ virtual asteroids sampled from the initial distribution of orbital uncertainty $q(\myvec{x})$ confirms the existence of a single VI, with an uncertainty of several days in the time of close approach. The LOV presents a sharp bend close to the center of the Earth and two branches. The initial MC exploration finds close approaches on both branches and the extended filter converges to several VIs on both the upper branch and the lower branch, which must be merged together. Only one VI per branch is shown in the figure. The circles and crosses represent the modified target plane coordinates of the set of virtual asteroids that are propagated numerically for the nonlinearity test. Several virtual asteroids from the upper-branch VI make it to the lower branch proving that both branches are dynamically connected. The dynamics along each branch is linear and the samples from the upper-branch VI are easily connected to the samples from the lower-branch VI using the criterion based on the partial derivatives of the modified target plane coordinates. The algorithm successfully merges all the VIs and reports a single VI.

\subsection{Generic completeness\label{Sec:completeness}}

Impact-monitoring systems are usually designed to find VIs above a minimum impact probability threshold, which is set by operational requirements. %\citet{milani2005nonlinear} and~\citet{del2019completeness} defined the completeness level of an impact-monitoring system as the highest impact probability of a VI that may escape detection. 
In order to detect a specific VI with the proposed algorithm, at least one of the samples propagated during the initial MC simulation with $N=10^4$ virtual asteroids must make a close approach within $d_\text{ca}=0.1~\text{au}$ along the same dynamical path. Then, starting from that initial guess, the extended filter converges to the corresponding impact trajectory assuming that it is compatible with the observational data. Therefore, the completeness of Sentry-II is driven by the probability of finding at least one close approach that is dynamically connected to a specific VI.

To quantify this probability we assume that, locally, the virtual asteroids are distributed along a straight line on the $b$-plane \citep{valsecchi2003resonant,milani2005nonlinear}. Given the impact probability of the VI under consideration, $P(F_i)$, the probability of finding a dynamically connected close approach at $d_\text{ca}$ can be obtained by scaling the probability of impact with the ratio $d_\text{ca}/\lambda R_\oplus$, where $\lambda$ accounts for the gravitational focusing during the close encounter and is defined in Eq.~\eqref{Eq:bplane_cond}. Thus, the expected number of close approaches found during the initial MC simulation is
\begin{equation}\label{Eq:expected_nca}
	n_\text{ca} = N P(F_i)\frac{d_\text{ca}}{\lambda R_\oplus}.
\end{equation}
\citet{milani2005nonlinear} and~\citet{del2019completeness} assume $\lambda=2$ when estimating the completeness of CLOMON2, which corresponds to $v_\infty=6.5~\text{km/s}$. Sentry and Sentry-II use $d_\text{ca}=0.1~\text{au}$ while CLOMON2 uses $d_\text{ca}=0.2~\text{au}$. The choice of $d_\text{ca}$ is a trade-off between computational performance and completeness and its effect depends on the implementation of each specific system \citep{milani2005nonlinear}.

Based on Eq.~\eqref{Eq:expected_nca}, the probability of finding at least one close approach within $d_\text{ca}$ that is dynamically connected to a specific VI can be estimated using Poisson statistics. Let $P_\text{ca}$ denote this probability, which increases with the actual impact probability of the VI and the $v_\infty$ of the encounter:
\begin{equation}\label{Eq:Pca}
	P_\text{ca}(P(F_i),v_\infty) = 1 - \Bigg[1 - P(F_i)\frac{d_\text{ca}}{\lambda(v_\infty) R_\oplus}\Bigg]^N.
\end{equation}
Using this expression, we define completeness as
\begin{equation}\label{Eq:completeness}
	C(P(F_i)) = \int_{0}^\infty P_\text{ca}(P(F_i),v_\infty)f(v_\infty)\,\mathrm{d}v_\infty,
\end{equation}
where $f(v_\infty)$ is the p.d.f.\ of the $v_\infty$ distribution for Earth impactors given by \citet{chesley2019development}. Equation~\eqref{Eq:completeness} is the expected value of the probability of finding a VI with impact probability $P(F_i)$ accounting for the actual distribution of hyperbolic excess velocities.

\begin{figure}
	\centering
	\includegraphics[width=\linewidth]{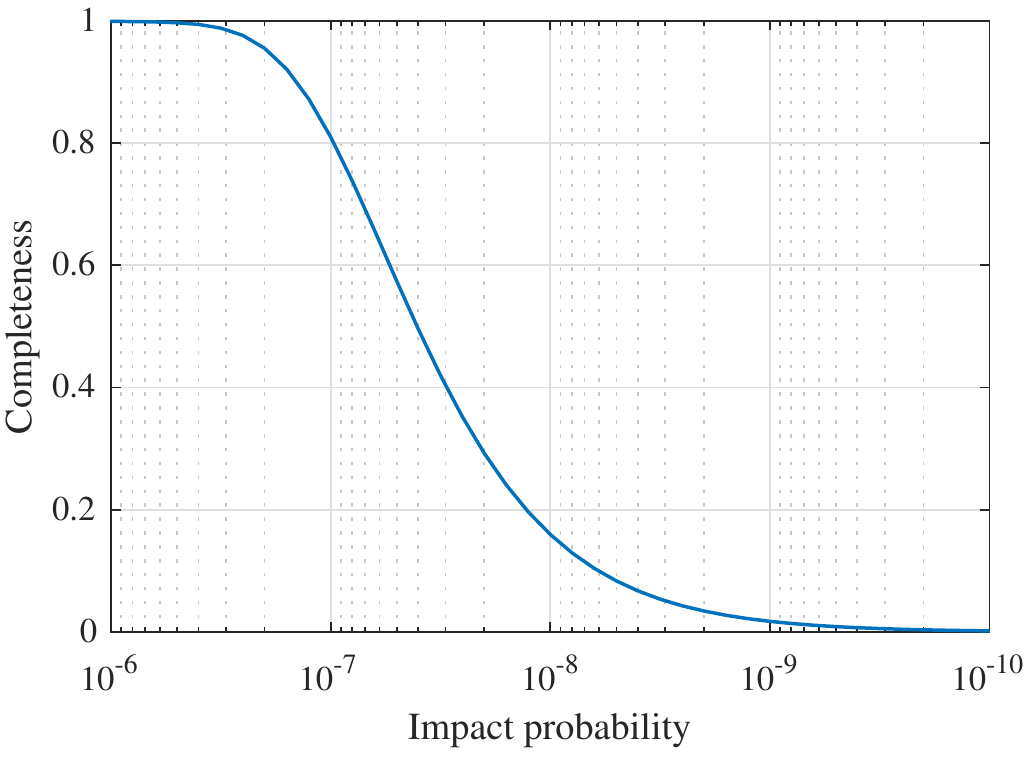}
	\caption{Completeness of Sentry-II for different impact probability levels.\label{Fig:completeness}}
\end{figure}

Figure~\ref{Fig:completeness} shows the completeness of the Sentry-II system as a function of the impact probability computed from Eq.~\eqref{Eq:completeness} using Eq.~\eqref{Eq:Pca}. The impact probability for 99\% completeness is $3\times10^{-7}$ and the system is approximately 80\% complete for $P(F_i)>10^{-7}$. The completeness for impact probabilities down to $10^{-8}$ is less than 20\%.

\subsection{RMS scaling for non-Gaussian initial uncertainty distributions\label{Sec:rms}}

Equation~\eqref{Eq:ip_definition} defines the impact probability in terms of the initial distribution of orbital uncertainty, $q(\myvec{x})$. Although in most cases assuming that this distribution is Gaussian provides a good approximation, there are cases when this assumption might not hold. This is particularly true for orbits determined from very short arcs whose uncertainty regions are relatively large and not likely to be accurately represented by a linear uncertainty distribution.

If the true distribution of uncertainty is given by a non-Gaussian distribution $r(\myvec{x})$---which is unknown---instead of by the normal distribution $q(\myvec{x})$, the probability estimate in Eq.~\eqref{Eq:ip_definition} transforms into:
\begin{equation}\label{Eq:ip_r}
    P(F_i) = \int_{F_i} r(\myvec{x})\,\diff \myvec{x}.
\end{equation}
Without loss of generality, assume that
\begin{equation}\label{Eq:r_proposal}
   r(\myvec{x}) \propto \exp(-\Delta\chi^2/2),
\end{equation}
where
\begin{equation}
    \Delta\chi^2 = Q - \bar{Q}
\end{equation}
is computed from the residuals without any assumption of linearity and $\bar{Q}$ is computed from the residuals along the nominal solution. Note that, if the uncertainty distribution is linear, then 
\begin{alignat}{1}
    \Delta \chi^2 &= Q - \bar{Q} = (\myvec{e}-\bar{\myvec{e}})^\top\mymatrix{W}(\myvec{e} - \bar{\myvec{e}}) \nonumber \\
    &\approx \Delta\myvec{x}^\top\mymatrix{B}^\top\mymatrix{W}\mymatrix{B}\Delta\myvec{x} = \Delta\myvec{x}^\top\mymatrix{\Sigma}^{-1}\Delta\myvec{x} = \sigma^2,
\end{alignat}
which means that Eq.~\eqref{Eq:r_proposal} becomes Gaussian.

Equation~\eqref{Eq:ip_r} is impractical because $r(\myvec{x})$ is not known and it cannot be sampled directly. Since $r(\myvec{x})$ is a p.d.f., it satisfies
\begin{equation}
    1 = \int r(\myvec{x})\,\diff\myvec{x} = \int \frac{r(\myvec{x})}{q(\myvec{x})}{q(\myvec{x})}\,\diff\myvec{x}.\label{Eq:expect_1}
\end{equation}
Assuming $r\propto\exp(-\Delta\chi^2/2)$, Eq.~\eqref{Eq:expect_1} can be approximated as
\begin{equation}\label{Eq:define_lambda}
    1\approx \frac{\gamma}{N_r}\sum_{k=1}^{N_r} \exp\bigg[{-\frac{1}{2}(\Delta\chi_k^2 - \sigma_{k}^2)}\bigg] ,\quad \myvec{x} \sim q(\myvec{x}),
\end{equation}
where $\gamma$ is a constant factor, $\Delta\chi^2_k=Q_k-\bar{Q}$, $\sigma_{k}^2=\{\Delta\myvec{x}\}_k^\top\mymatrix{\Sigma}^{-1}\{\Delta\myvec{x}\}_k$, and we set $N_r=10^3$ in Sentry-II. Equation~\eqref{Eq:define_lambda} can be solved for the scaling factor $\gamma$, resulting in:
\begin{equation}
    \gamma \approx \frac{N_r}{ \sum_{k=1}^{N_r} \exp\bigg[{-\frac{1}{2}(\Delta\chi_k^2 - \sigma_{k}^2)}\bigg] }.
\end{equation}
Finally, the probability estimate in Eq.~\eqref{Eq:ip_is} is corrected using the RMS scaling:
\begin{alignat}{1}
    P(F_i) &= \int_{F_i} r(\myvec{x})\,\diff \myvec{x} = \int_{F_i} \frac{r(\myvec{x})}{q(\myvec{x})}\frac{q(\myvec{x})}{p_i(\myvec{x})}p_i(\myvec{x})\,\diff \myvec{x} \nonumber \\
    %&\approx \frac{r(\bar{\myvec{x}}_{i})}{q(\bar{\myvec{x}}_{i})}\int_{F_i} \frac{q(\myvec{x})}{p_i(\myvec{x})}p_i(\myvec{x})\,\diff \myvec{x} \nonumber\\
    & \approx \frac{1}{N_{\text{ip}}}\sum_{k=1}^{N_{\text{ip}}} I_{F_i}(\{\myvec{x}\}_k)\frac{r(\{\myvec{x}\}_k)}{q(\{\myvec{x}\}_k)}\frac{q(\{\myvec{x}\}_k)}{p_i(\{\myvec{x}\}_k)}\nonumber\\
    & \approx \frac{1}{N_{\text{ip}}}\frac{r(\bar{\myvec{x}}_{i})}{q(\bar{\myvec{x}}_{i})}\sum_{k=1}^{N_{\text{ip}}} I_{F_i}(\{\myvec{x}\}_k)w_{i,k}.
\end{alignat}
The ratio $r/q$ is approximated using its value along the nominal trajectory of the VI to avoid the computation of $\Delta\chi^2$ for each of the virtual asteroids. Equation~\eqref{Eq:define_lambda} furnishes the adjusted probability estimate:
\begin{alignat}{1}
    \tilde{P}(F_i) \approx & \frac{\gamma}{N_{\text{ip}}}\exp\bigg[{-\frac{1}{2}(\Delta\chi_i^2 - \sigma_{i}^2)}\bigg] \sum_{k=1}^{N_{\text{ip}}}[ I_{F_i}(\{\myvec{x}\}_k)w_{i,k}].\label{Eq:prob_rms}
\end{alignat}
The variance of the adjusted probability estimate is:
\begin{equation}
    \tilde{\sigma}_{P_i}^2 =   \frac{1}{N_\text{ip}} \Bigg\{ \frac{\gamma^2}{N_\text{ip}}\exp(\sigma_{i}^2-\Delta\chi_i^2)\sum_{k=1}^{N_\text{ip}} \big[I_{F_i}(\{\myvec{x}\}_k)w_{i,k}^2\big] - \big[\tilde{P}(F_i)\big]^2 \Bigg\}.
\end{equation}

If the initial distribution of orbital uncertainty is truly linear, then $\gamma\to1$, $\Delta\chi_i^2\to\sigma_i^2$, and Eq.~\eqref{Eq:prob_rms} reduces to Eq.~\eqref{Eq:ip_is}. The values that $\gamma$ and the exponential factor in Eq.~\eqref{Eq:prob_rms} take can be used to assess how well the assumption of linearity of the initial distribution of orbital uncertainty holds.

\section{Validation}\label{Sec:examples}
%To evaluate the performance of the method, we processed all known NEAs searching for VIs compatible with their distribution of orbit uncertainty and estimated their impact probability. 

We conducted an extensive campaign to test the proposed impact monitoring algorithm. Sentry-II has processed all known NEAs searching for VIs compatible with their orbital uncertainty distribution. In this section, we first discuss selected cases of increasing complexity to show how the method works in practice. Next, we compare the results with the results obtained with direct MC simulation and the LOV method, particularly with the Sentry system.

\subsection{Selected examples}

\subsubsection{2020~WB$_3$ (orbit solution JPL~2)}
The first close approach of asteroid 2020~WB$_3$ occurs in 2024, close to its next perihelion passage. The orbit fit uses 42~observations between 2020-Nov-21 and 2020-Nov-23. The extended filter incorporating the impact pseudo-observation converges to an impact trajectory with nominal impact date 2024-Sep-28.75 and impact probability $P=3.5\times10^{-7}$. The asteroid completes less than one revolution and the dynamics remains linear along the trajectory leading to the impact.

\begin{figure}[h]
	\centering
	\includegraphics[width=\linewidth]{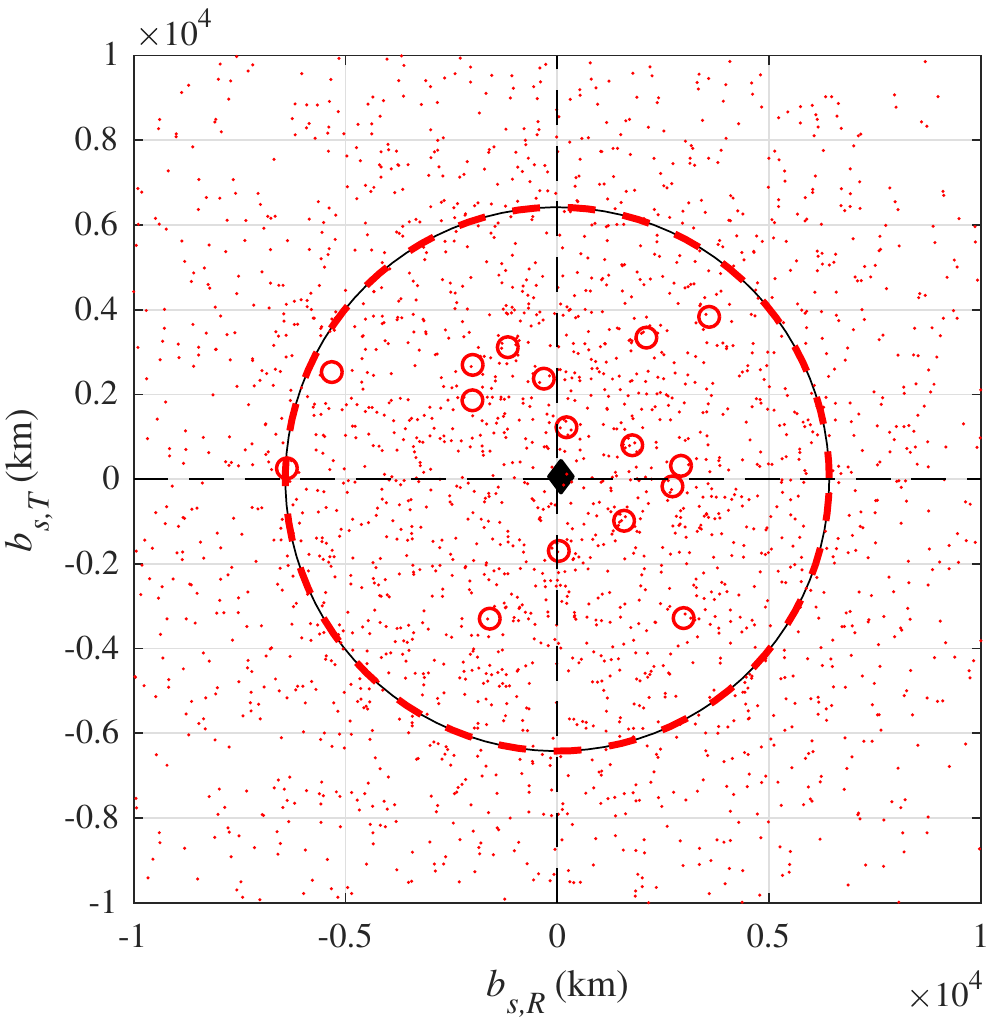}
	\caption{Projection on the scaled $b$-plane of the 1-sigma covariance ellipsoid of the 2024-Sep-28.75 VI of asteroid 2020~WB$_3$ (red dashed line). The black circumference represents the Earth's cross section, the black diamond is the nominal impact location, the dots are the virtual asteroids propagated with the linear mapping for importance sampling, and the circular markers are the virtual asteroids propagated numerically during the nonlinearity test (only the impacting ones are shown).\label{Fig:example_2020wb3}}
\end{figure}

Figure~\ref{Fig:example_2020wb3} shows the 1-sigma covariance ellipse of the VI mapped to the scaled $b$-plane. After such a short propagation (less than one orbital revolution), the stretching of the uncertainty region is negligible and the projection of the covariance ellipsoid on the scaled $b$-plane coincides with the Earth cross section. The dots represent the $N_\text{ip}$ samples mapped linearly with Eq.~\eqref{Eq:linear_bmap} to estimate the impact probability using importance sampling. The circular markers are the virtual asteroids that are propagated numerically during the nonlinearity test and impact Earth. Recall that the VI covariance is scaled down by a factor of two when sampling these virtual asteroids. In this example, the nonlinear samples are well approximated with the linear mapping. The black diamond represents the scaled $b$-plane coordinates of the nominal VI trajectory.

\subsubsection{2008~TS$_{10}$ (orbit solution JPL~7)}

\begin{figure}[h]
	\centering
	\includegraphics[width=\linewidth]{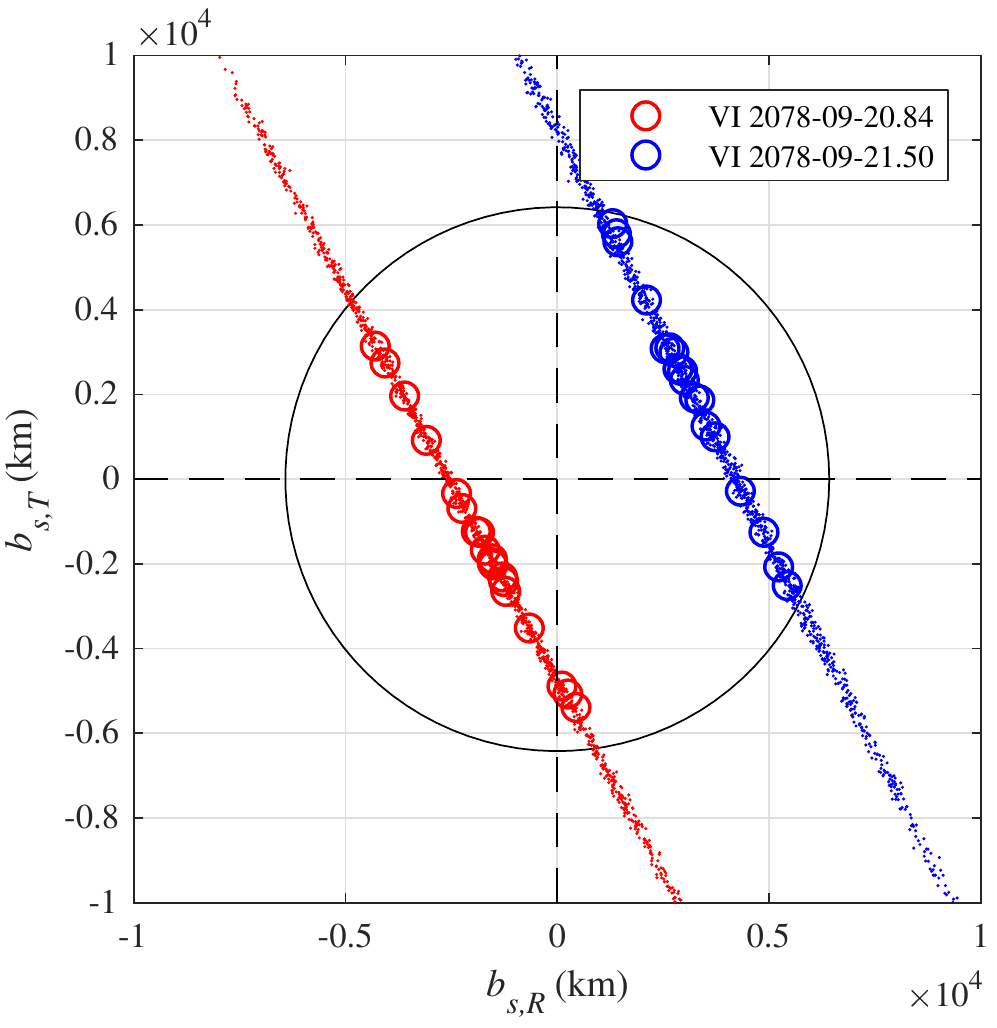}
	\caption{Projection on the scaled $b$-plane of the virtual asteroids used for importance sampling (propagated using the linear mapping and represented by dots) and for the nonlinearity test (propagated numerically and represented by circular markers) for the September 2078 VIs of asteroid 2008~TS$_{10}$.\label{Fig:example_2008ts10}}
\end{figure}

Asteroid 2008~TS$_{10}$ admits two VIs in September 2078, with similar nominal impact dates. The orbit fit uses 17~observations spanning 2008-Oct-08 through 2008-Oct-22. The extended filter identified these two separate VIs thanks to starting from initial conditions on each dynamical path. The projections on the scaled $b$-plane are shown in Figure~\ref{Fig:example_2008ts10}. Both VIs are assumed linear based on the nonlinearity test and the impact probability is estimated via importance sampling mapping the virtual asteroids to the scaled $b$-plane using the linear approximation. The impact probability of both VIs is $3.5\times10^{-6}$.

%\subsubsection{2020~XX$_3$ (orbit solution JPL~2)}
%Four different VIs with nominal impact date 2088-Dec-17 were found when exploring the orbital uncertainty distribution of asteroid 2020~XX$_3$. Figure~\ref{Fig:example_2020xx3} shows their projection on the scaled $b$-plane. This case is particularly interesting because the nominal impact location of the 2088-Dec-17.50 VI is very close to the edge of the Earth's cross-section. The VI covariance is produced with the scaling in Eq.~\eqref{Eq:sigma_b} using $\alpha=1$, which results in an impact ratio 

%\begin{figure}[h]
%	\centering
%	\includegraphics[width=\linewidth]{figures/2020xx3_ca2088.pdf}
%	\caption{Projection on the scaled $b$-plane of the 1-sigma covariance ellipsoid of the 2088-Dec-17 VIs of asteroid 2020~XX$_{3}$.\label{Fig:example_2020xx3}}
%\end{figure}

\subsubsection{2017~LD (orbit solution JPL~15)}

\begin{figure}[h]
	\centering
	\includegraphics[width=\linewidth]{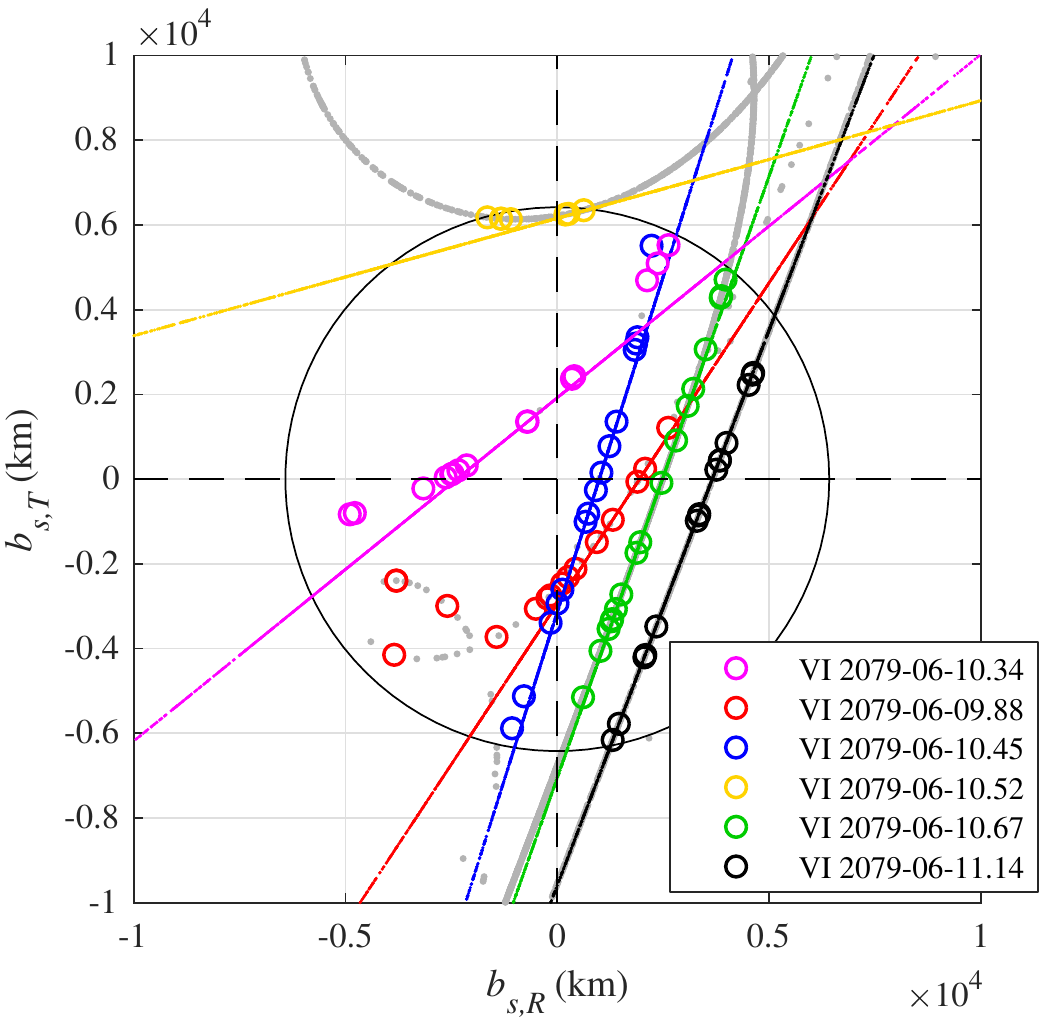}
	\caption{Projection on the scaled $b$-plane of the virtual asteroids used for importance sampling (propagated using the linear mapping and represented by dots) and for the nonlinearity test (propagated numerically and represented by circular markers) for the June 2079 VIs of asteroid 2017~LD. Gray dots correspond to a set of $10^7$ MC samples propagated numerically.\label{Fig:example_2017ld}}
\end{figure}

Figure~\ref{Fig:example_2017ld} shows the projections on the scaled $b$-plane of the six VIs of asteroid 2017~LD corresponding to an impact in June, 2079. The orbit fit uses 133~observations between 2017-May-16 and 2017-Jun-30. A set of $10^7$ MC samples propagated numerically to the scaled $b$-plane (represented with gray dots) depicts the LOV and shows the strongly nonlinear character of the problem. 

Table~\ref{Tab:prob_2017ld} presents the impact probability of each VI. There are three linear VIs according to the nonlinearity test: 2079-Jun-10.45, 2079-Jun-10.67, and 2079-Jun-11.14. It is worth noting that the VI on 2079-Jun-10.67 is linear within the Earth cross section---the only region of interest---although the LOV curves outside of the Earth.

The remaining three VIs are nonlinear, as shown by the curvature of the LOV inside the Earth cross section. The curvature is captured by the samples propagated numerically during the nonlinearity test. For these three cases, instead of using the linear approximation to propagate the virtual asteroids for importance sampling, Sentry-II resorts to an MC simulation with 200~virtual asteroids. These samples are not shown in Fig.~\ref{Fig:example_2017ld}. The linear approximation would have underestimated the impact probability of the VI on 2079-Jun-09.88 by a factor of two and would have overestimated that of the VI on 2079-Jun-10.52 by more than one order of magnitude. The impact probability of the VI on 2079-Jun-10.34 is too low for the $10^7$ MC samples to represent the LOV accurately.

\begin{table}
	\caption{Nominal impact date and impact probability of the June, 2079, VIs of asteroid 2017~LD.\label{Tab:prob_2017ld}}
	\centering
	\begin{tabular}{ccc}
	\hline\hline\noalign{\smallskip}
	Impact date & $P(F)$ & Linear \\
\noalign{\smallskip}\hline\noalign{\smallskip}
2079-Jun-09.88 & $2.1\times10^{-5}$ & N \\ % linear 1.2E-5
2079-Jun-10.34 & $6.0\times10^{-7}$ & N \\ % linear 5.4E-7
2079-Jun-10.45 & $1.4\times10^{-6}$ & Y \\
2079-Jun-10.52 & $2.2\times10^{-7}$ & N \\ % linear 8.1E-6
2079-Jun-10.67 & $5.1\times10^{-4}$ & Y \\
2079-Jun-11.14 & $1.8\times10^{-4}$ & Y \\
	\bottomrule
	\end{tabular}
\end{table}

\subsubsection{2000~SB$_{45}$ (orbit solution JPL~11)}

\begin{figure}[h]
	\centering
	\includegraphics[width=\linewidth]{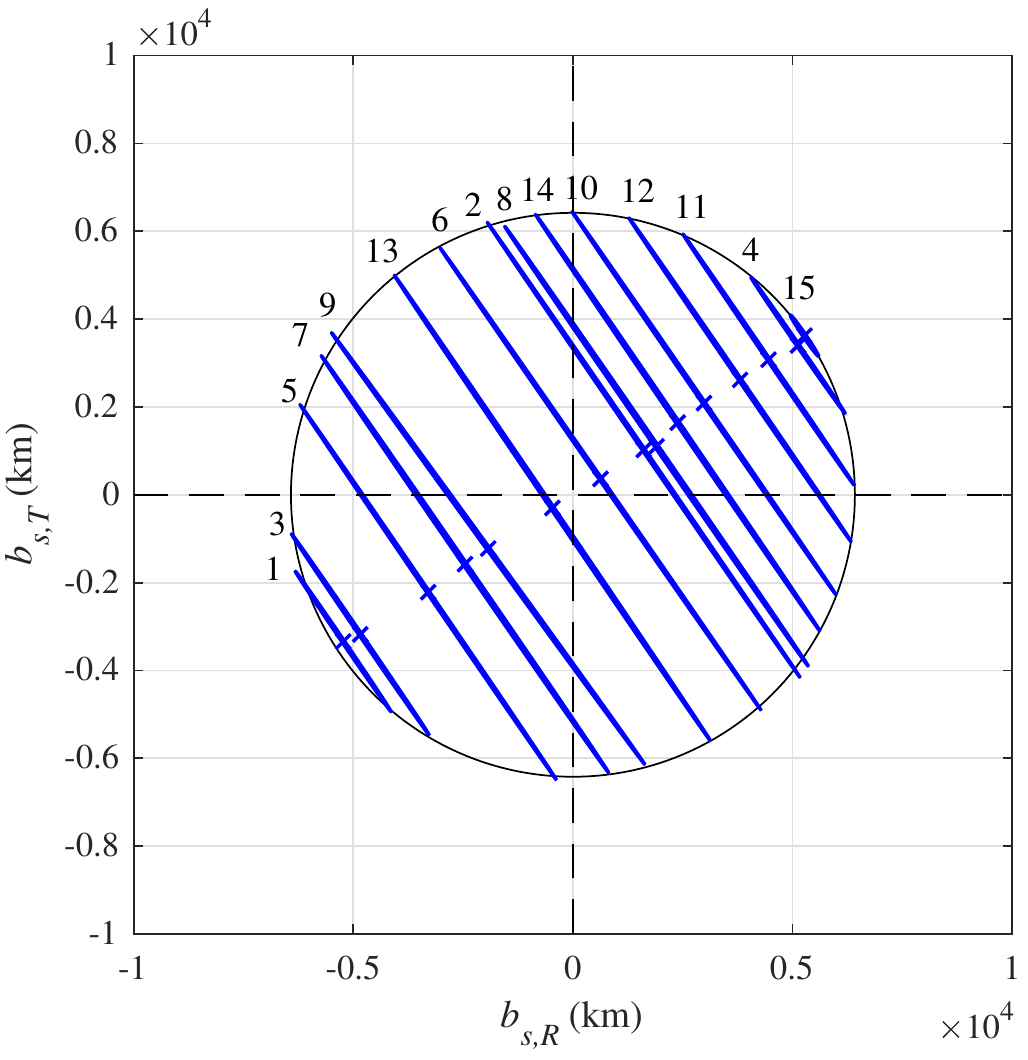}
	\caption{Projections on the scaled $b$-plane of the 1-sigma covariance ellipsoid of the October 2084 VIs of asteroid 2000~SB$_{45}$. Crosses represent the nominal impact points.\label{Fig:example_2000bs45}}
\end{figure}

The orbital uncertainty distribution of asteroid 2000~SB$_{45}$ (resulting from fitting 18~observations between 2000-Sep-27 and 2000-Sep-29) is compatible with tens of VIs in the next 100~years. This is a challenging case because there are a large number of distinct dynamical paths leading to impacts on specific dates. For example, Fig.~\ref{Fig:example_2000bs45} depicts 15~VIs with nominal impact date 2084-Oct-08. The new system is able to find the VIs, determine that they are not dynamically connected, and estimate their impact probability. Note how the system scales down the uncertainty of the impact pseudo-observation to compensate for the VIs being close to the edge of the Earth's disk. 

Table~\ref{Tab:prob_2000sb45} presents the impact probability of the VIs shown in Fig.~\ref{Fig:example_2000bs45}. The cumulative impact probability is $2.6\times10^{-5}$, which is in good agreement with the MC ($2.6\times10^{-5}$) and Sentry ($2.5\times10^{-5}$) estimates. The virtual asteroids used for importance sampling are propagated to the scaled $b$-plane using the linear mapping because none of the VIs are flagged as nonlinear.

\begin{table}
	\caption{Nominal impact date and impact probability of the October 8, 2084, VIs of asteroid 2000~SB$_{45}$. Id numbers are used to identify each VI in Fig.~\ref{Fig:example_2000bs45}.\label{Tab:prob_2000sb45}}
	\centering
	\begin{tabular}{ccc}
	\hline\hline\noalign{\smallskip}
	Id & Impact date & $P(F)$ \\
\noalign{\smallskip}\hline\noalign{\smallskip}
1 & 2084-Oct-08.02 & $1.6\times10^{-9}$ \\
2 & 2084-Oct-08.07 & $2.7\times10^{-8}$ \\
3 & 2084-Oct-08.08 & $5.3\times10^{-9}$ \\
4 & 2084-Oct-08.08 & $5.3\times10^{-9}$ \\
5 & 2084-Oct-08.10 & $4.3\times10^{-9}$ \\
6 & 2084-Oct-08.12 & $4.3\times10^{-8}$ \\
7 & 2084-Oct-08.13 & $8.7\times10^{-8}$ \\
8 & 2084-Oct-08.13 & $2.5\times10^{-8}$ \\
9 & 2084-Oct-08.16 & $1.3\times10^{-8}$ \\
10 & 2084-Oct-08.17 & $1.4\times10^{-7}$ \\
11 & 2084-Oct-08.18 & $4.3\times10^{-8}$ \\
12 & 2084-Oct-08.20 & $2.2\times10^{-7}$ \\
13 & 2084-Oct-08.21 & $2.3\times10^{-5}$ \\
14 & 2084-Oct-08.23 & $2.0\times10^{-6}$ \\
15 & 2084-Oct-08.23 & $4.8\times10^{-8}$ \\
	\bottomrule
	\end{tabular}
\end{table}

\subsubsection{(410777) 2009~FD (orbit solution JPL~100)}

The 2190 VI of asteroid (410777) 2009~FD is driven by the Yarkovsky effect and a scattering encounter in 2185. The perturbation due to the Yarkovsky effect is modeled as a transversal acceleration
\begin{equation}
	\myvec{a}_p = \frac{A_2}{(r / 1~\text{au})^2}\,\myvec{t},
\end{equation}
written in terms of a dynamical parameter $A_2$, the heliocentric radius $r$, and the unit vector pointing in the transversal direction, $\myvec{t}$ \citep{bottke2006yarkovsky,chesley2015direct,farnocchia2013near,farnocchia2013yarkovsky}. In practice, $A_2$ is appended to the vector of parameters $\myvec{x}$ and estimated as part the orbit-determination process.

\begin{figure}[h]
	\centering
	\includegraphics[width=\linewidth]{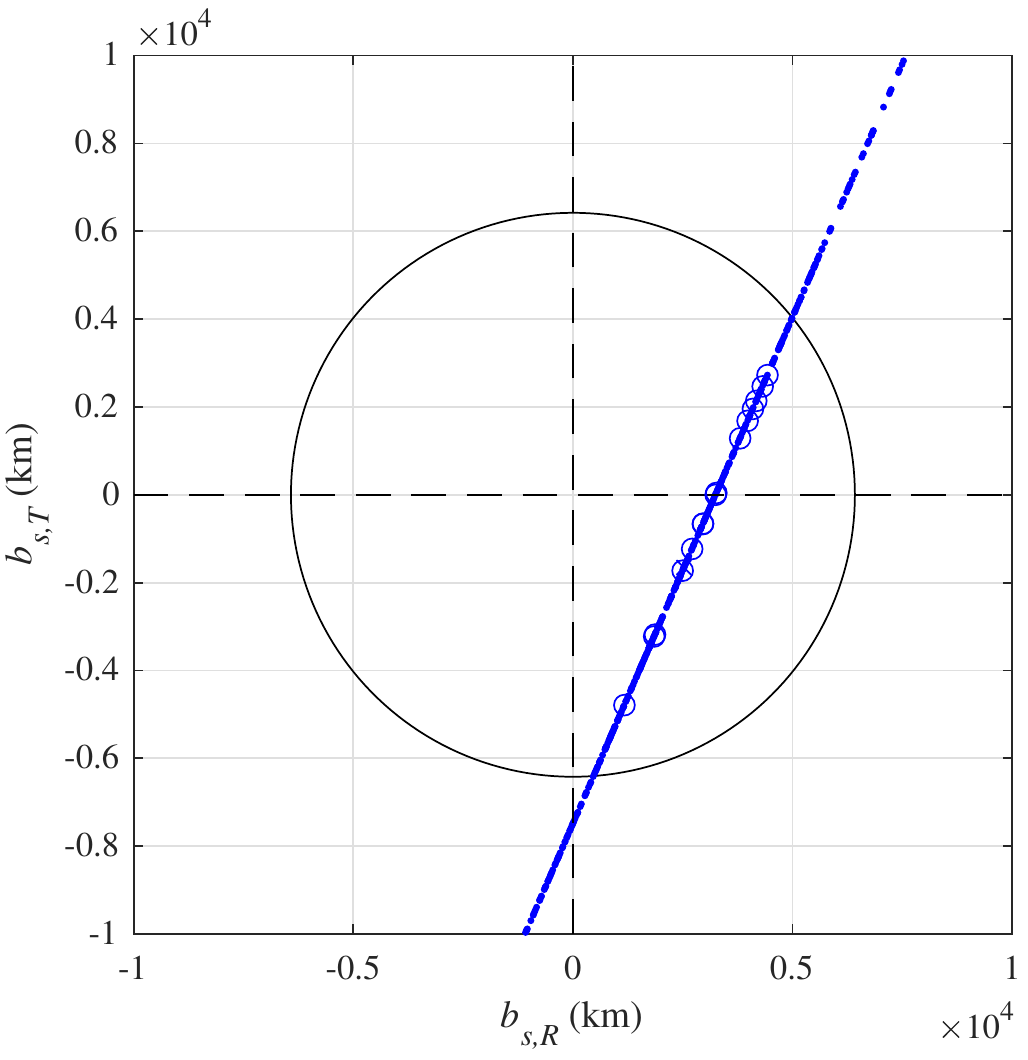}
	\caption{Projections on the scaled $b$-plane of the virtual asteroids sampled from the uncertainty distribution of the 2190-Mar-30.08 VI of asteroid (410777) 2009~FD propagated with the partial derivatives (dots, only 500~samples are shown) and numerically (circular markers, only the impacting solutions are shown).\label{Fig:example_a410777}}
\end{figure}

\citet{vigna2019yarkovsky} published a dedicated analysis of the 2190 VI to properly explore the region of orbital uncertainty including $A_2$. The nominal orbit solution relies on 497~observations (including 5~radar delay and 3~radar Doppler) spanning 2009-Feb-24 through 2019-Apr-15. The analysis was based on a special definition of the LOV \citep{spoto2014nongravitational} and a sequential MC algorithm \citep{roa2019multilayer}. The algorithm presented in this paper can robustly handle the analysis of this VI without manual intervention thanks to being transparent to the nature of the parameters included in $\myvec{x}$. The impact probability estimate from Sentry-II is $(1.1\pm0.2)\times10^{-8}$, which is consistent with the MC estimate of $(1.5\pm0.4)\times10^{-8}$ from \citet{vigna2019yarkovsky}. Figure~\ref{Fig:example_a410777} represents the scaled $b$-plane during the 2190 close approach, revealing the linear nature of the dynamics.

The extended filter converges to the nominal orbit of the VI and provides the covariance of the fit, which approximates the region in parameter space where impacts may be found. As a result, the filter provides the nominal value and uncertainty of $A_2$ for the VI, $A_2=(-5.9\pm0.5)\times10^{-15}~\text{au}/\text{d}^{2}$, which is consistent with the interval of values leading to impacts using MC reported in \citet{vigna2019yarkovsky}.

\subsection{Validation with MC sampling and Sentry}
Table~\ref{Tab:compare_MC} compares the cumulative impact probability estimates for multiple asteroids obtained with the new system, an MC simulation using $10^7$ virtual asteroids, and the LOV method implemented in Sentry. The time span is limited to 100~years into the future. The asteroids in the table are among the top entries of the Sentry risk list.\footnote{https://cneos.jpl.nasa.gov/sentry/}

The probability estimates match within 3-sigma for most asteroids. The first exception is asteroid 2012~HG$_2$. The orbital inclination of this asteroid above the ecliptic is negligible, it undergoes slow encounters with Earth at perihelion and with Mars at aphelion with $v_\infty<5~\text{km/s}$, and it is close to a 13:1 resonance with Earth. Hundreds of VIs are found for this asteroid. These features make it dynamically challenging and numerically sensitive. Sentry overestimates the impact probability of the dominant VIs in 2071 and 2078 by approximately a factor of two, resulting in a higher cumulative probability compared to MC sampling. Tightening the tolerances of the orbit-determination program allows the new system to overcome numerical instabilities and match the MC result. The next exception is 2000~SG$_{344}$. Its orbital period of almost one year, near-zero inclination above the ecliptic, and apoapsis just above Earth's orbit result in very low encounter velocities ($v_\infty=1.4~\text{km/s}$) and a continuum of close approaches after 2070, as it was previously shown in Fig.~\ref{Fig:ca_dates}. The special dynamics of the asteroid has motivated several dedicated analyses of its impact hazard \citep{chodas20012000,chesley2002quantifying}. Both Sentry and Sentry-II provide lower cumulative probability estimates compared to MC simulation. For the case of Sentry-II, the system underestimates by 10--20\% the cumulative impact probability the VIs after 2102. The problem becomes nonlinear and numerically sensitive, affecting the accuracy of the propagation of the variational equations and the stability of the algorithm for estimating the impact probability. This issue can be partially alleviated by forcing Sentry-II to use MC simulation for importance sampling when estimating the impact probability of the problematic VIs.

\begin{table}
\caption{Cumulative impact probability estimates from Sentry-II compared to MC simulation and Sentry.\label{Tab:compare_MC}}
\centering
\small
\begin{tabular*}{\linewidth}{@{\extracolsep{\fill}}lcccc@{}}
%\begin{tabular}{lllll}
\hline\hline\noalign{\smallskip}
\multicolumn{1}{c}{Asteroid} & \multicolumn{1}{c}{MC} & \multicolumn{1}{c}{$\sigma_\text{MC}$} & \multicolumn{1}{c}{Sentry-II} & \multicolumn{1}{c}{Sentry} \\
\noalign{\smallskip}\hline\noalign{\smallskip}
%
%2017~LD & $(2.29 \pm 0.02)\times10^{-3}$ & $2.25\times10^{-3}$ \\
%2000~SB$_{45}$ & $(1.53 \pm 0.04)\times10^{-4}$ & $1.59\times10^{-4}$ \\
%2009~JF$_{1}$ & $(2.53 \pm 0.05)\times10^{-4}$ & $2.59\times10^{-4}$ \\
%2005~QK$_{76}$ & $(7.04 \pm 0.27)\times10^{-5}$ & $6.80\times10^{-5}$ \\
%2010~RF$_{12}$  \\
%
%
2010 RF$_{12}$ & $4.8\times10^{-2}$ & $6.7\times10^{-5}$ & $4.8\times10^{-2}$ & $4.7\times10^{-2}$ \\
2017 WT$_{28}$ & $1.1\times10^{-2}$ & $3.3\times10^{-5}$ & $1.1\times10^{-2}$ & $1.1\times10^{-2}$ \\
2020 VW    & $7.0\times10^{-3}$ & $3.9\times10^{-5}$ & $7.1\times10^{-3}$ & $7.0\times10^{-3}$ \\
2012 HG$_2$    & $1.7\times10^{-3}$ & $1.3\times10^{-5}$ & $1.7\times10^{-3}$ & $2.8\times10^{-3}$ \\
2000 SG$_{344}$& $3.3\times10^{-3}$ & $1.8\times10^{-5}$ & $2.9\times10^{-3}$ & $2.6\times10^{-3}$ \\
2020 VV    & $2.5\times10^{-3}$ & $1.6\times10^{-5}$ & $2.3\times10^{-3}$ & $2.3\times10^{-3}$ \\
2013 VW$_{13}$ & $4.4\times10^{-4}$ & $6.7\times10^{-6}$ & $4.3\times10^{-4}$ & $4.3\times10^{-4}$ \\
2009 JF$_1$    & $2.6\times10^{-4}$ & $5.1\times10^{-6}$ & $2.6\times10^{-4}$ & $2.6\times10^{-4}$ \\ % OLD
2008 JL$_3$    & $1.6\times10^{-4}$ & $4.0\times10^{-6}$ & $1.6\times10^{-4}$ & $1.6\times10^{-4}$ \\
2000 SB$_{45}$ & $1.8\times10^{-4}$ & $4.3\times10^{-6}$ & $1.6\times10^{-4}$ & $1.6\times10^{-4}$ \\ % OLD
1994 GK        & $7.2\times10^{-5}$ & $2.7\times10^{-6}$ & $6.9\times10^{-5}$ & $6.9\times10^{-5}$ \\
2005 QK$_{76}$ & $6.6\times10^{-5}$ & $2.6\times10^{-6}$ & $6.8\times10^{-5}$ & $6.8\times10^{-5}$ \\ % OLD
2007 DX$_{40}$ & $7.5\times10^{-5}$ & $2.7\times10^{-6}$ & $6.5\times10^{-5}$ & $6.2\times10^{-5}$ \\
2008 EX$_5$    & $4.9\times10^{-5}$ & $2.2\times10^{-6}$ & $4.8\times10^{-5}$ & $4.7\times10^{-5}$ \\
2008 UB$_7$    & $4.1\times10^{-5}$ & $2.0\times10^{-6}$ & $3.7\times10^{-5}$ & $3.5\times10^{-5}$ \\
\bottomrule
\end{tabular*}
%\end{tabular}
\end{table}

\subsection{Statistical comparison with Sentry}
\begin{figure}
	\centering
	\gridline{\fig{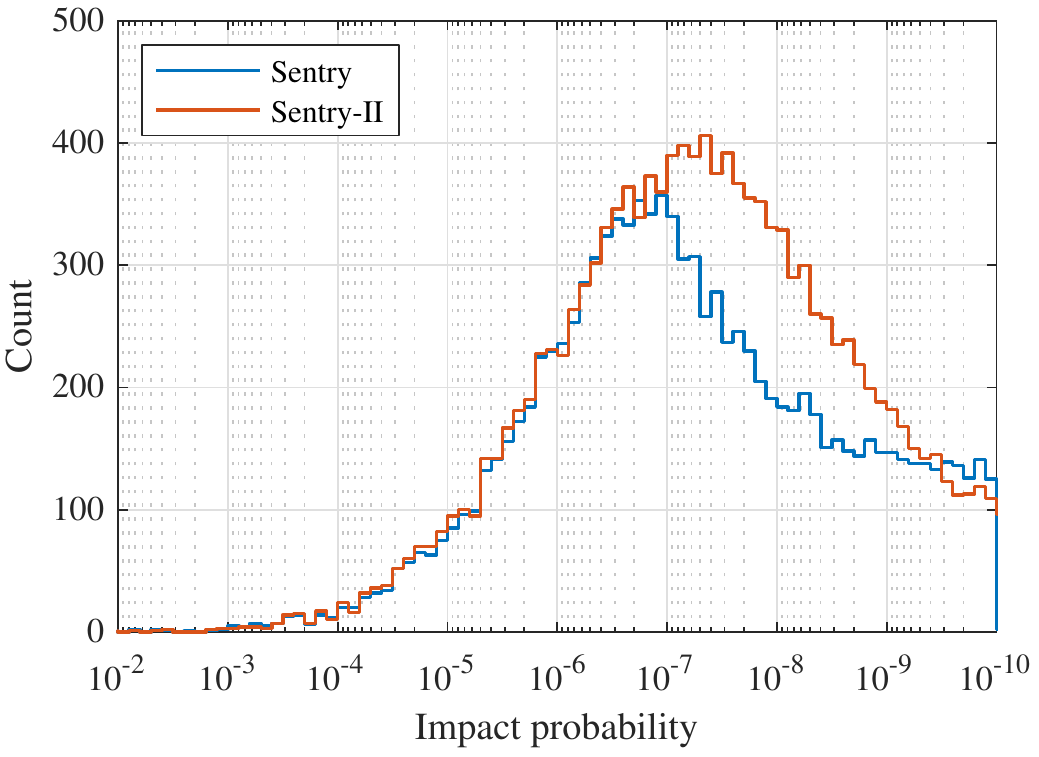}{\linewidth}{(a) Number of asteroids with at least one VI within each impact probability bin}}
	\gridline{\fig{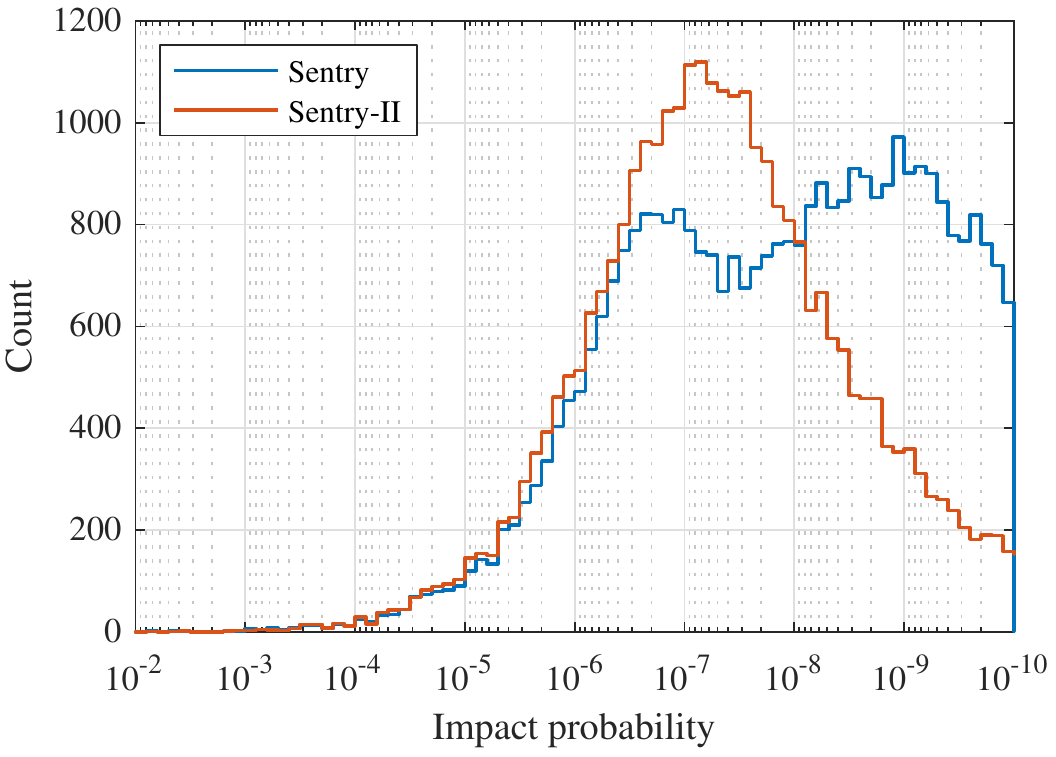}{\linewidth}{(b) Number of distinct VI impact dates within each impact probability bin}}
	\caption{Comparison of the number of VIs found by Sentry and Sentry-II between 2021-Jan-01 and 2121-Jan-01.\label{Fig:histogram_sentry}}
\end{figure}

We processed all known NEAs with the new Sentry-II system and compared the results with Sentry in a statistical sense. First, we count the number of asteroids for which at least one VI was found between 2021-Jan-01 and 2121-Jan-01, using different impact probability intervals. Figure~\ref{Fig:histogram_sentry}a shows the resulting histograms for Sentry and the new system. Both systems provide very similar results for impact probabilities down to $10^{-7}$. The number of asteroids with VIs according to the new system keeps increasing until reaching an impact probability of approximately $4\times10^{-8}$. The new system found in total 1,214 asteroids with at least one VI with $P\geq10^{-10}$, which is a 7\% increase compared to the 1,134 asteroids reported by Sentry. The extended filter proves more robust for finding VIs within the completeness level discussed in Section~\ref{Sec:completeness}.

Next, we process the list of VIs for each asteroid given an impact probability interval and count the number of distinct impact dates. Figure~\ref{Fig:histogram_sentry}b presents histograms for the total number of impact dates found within each impact probability bin. The number of distinct impact dates reported by the new system and by Sentry match for impact probabilities greater than $4\times10^{-7}$. The new system reports more impact dates down to impact probabilities of $10^{-8}$. The distribution of VI dates from Sentry is bimodal, with an increase in the number of VI dates for impact probabilities of $10^{-9}$. The distribution of VIs found by Sentry-II is consistent with the completeness analysis by \citet{del2019completeness}.

 %The statistical results presented in this section suggest that, thanks to the multi-dimensional exploration of the initial distribution of orbital uncertainty and not relying on interpolating along the LOV, the proposed algorithm may be more robust when detecting VIs within its target completeness level.

%The results from the new system are consistent with Sentry and indicate that it may be more robust when detecting VIs within its target completeness level.

\section{Computational performance\label{Sec:runtime}}

The main motivation for developing the new impact monitoring technique presented in this paper is to be more robust and more general than the LOV method. Even if robustness is the main factor driving the design, it is still important that the runtime remains comparable to the LOV method for the proposed technique to be operationally attractive.

\begin{figure}
	\centering
	\includegraphics[width=\linewidth]{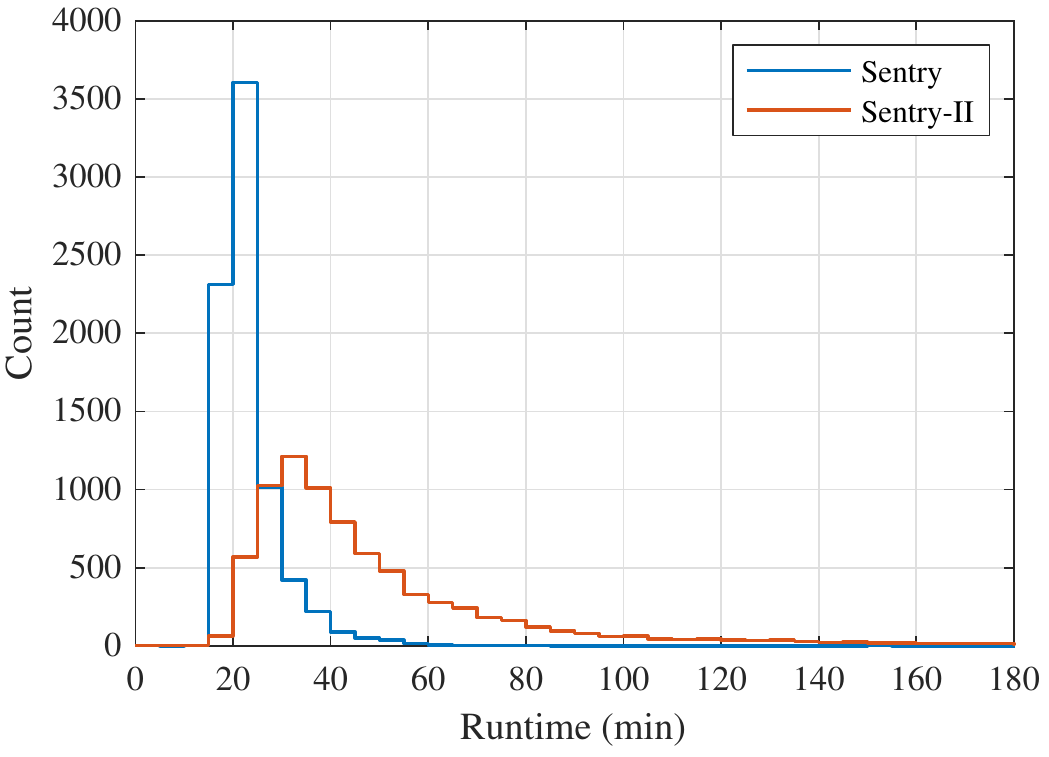}
	\caption{{Runtime comparison between Sentry and Sentry-II.}\label{Fig:runtime}}
\end{figure}

Figure~\ref{Fig:runtime} compares the distribution of runtimes for all the NEAs processed by both systems. The median runtime for Sentry-II is 41~minutes, which is approximately two times longer than Sentry's (21~minutes). The mean runtime for Sentry-II is 58~minutes, about {two times} longer than Sentry's mean runtime of 24~minutes. The peak of the runtime distribution of Sentry-II is at {30}~minutes, whereas the peak of Sentry's distribution is at 20~minutes. The maximum runtimes observed for Sentry-II and Sentry are {1.7} and 1.1~days, respectively.

A slowdown of {a factor of two}, with the median runtime remaining below {one hour}, is orders of magnitude faster than direct MC simulation and suitable for daily routine operations. For example, propagating $10^7$ virtual asteroids for 100~years using the same integration setup as Sentry and Sentry-II typically takes 14~days, which is {500 times} slower than the median runtime of Sentry-II.

\section{Conclusions}
Incorporating the impact condition as an observation into an orbit-determination program drives the filter toward impacting solutions, as long as they are compatible with the data. The observational data typically constrain the strong direction of the orbital uncertainty distribution, while the impact pseudo-observation constrains the weak direction. The extended filter iterates along the weak direction to minimize the close-approach distance.

One of the main advantages of the proposed technique for impact monitoring is that it explores the full multi-dimensional initial distribution of orbital uncertainty without simplifying assumptions about the dynamics. This feature allows the method to handle estimated parameters systematically without requiring manual intervention. Minimizing the need for human interaction will prove critical in the coming years when surveys such as LSST become operational and the number of NEAs to process grows dramatically. The results after processing thousands of NEAs suggest that the new technique can be more robust than LOV-based methods when detecting VIs given an impact probability threshold.

The extended filter returns the nominal VI orbit together with its uncertainty distribution, which is assumed to be Gaussian. The covariance ellipsoid captures the set of parameters that lead to an impact along that particular dynamical path and provides a convenient representation of the VI. Once the VI is fully characterized by a p.d.f.\ in parameter space, its impact probability can be estimated using standard variance-reduction techniques such as importance sampling. In addition, the uncertainty distribution of a VI can be propagated over time and projected onto the plane of sky to support negative observation campaigns to rule out VIs \citep{milani2000virtual}. As an example, a recent negative observation campaign successfully drove to zero the impact probability of asteroid 2006~QV$_{89}$.\footnote{https://www.eso.org/public/announcements/ann19039/ https://cneos.jpl.nasa.gov/news/news204.html}

Existing operational orbit-determination software can be extended to find impact trajectories by incorporating the residuals associated with the impact pseudo-observation. This is a notable feature of the proposed technique from an operational perspective. It allows development of an automatic impact-monitoring system relying on existing, thoroughly tested routines for orbit propagation, orbit determination, and data processing. The resulting impact-monitoring system inherits the capabilities of the parent orbit-determination program.

We implemented the proposed technique into a new automatic impact-monitoring system at JPL, called Sentry-II. The results presented in this paper are part of an extensive testing campaign to evaluate the robustness of Sentry-II and to optimize the selection of the configuration parameters. The system successfully processes asteroids with challenging dynamical features, like nongravitational forces and strong nonlinear phenomena. The results indicate that the proposed algorithm can discover multiple new VIs and provide reliable impact probability estimates when compared with existing impact-monitoring systems. The performance in terms of runtime is comparable to existing systems, with the Sentry-II median runtime of 41~minutes being approximately two times longer than Sentry's. We estimate that Sentry-II is 99\% complete for VIs with impact probabilities down to $3\times10^{-7}$. The completeness of Sentry-II is comparable to that of Sentry, which should be considered as a bound given the limitations of Sentry when dealing with pathological cases.

\section*{Acknowledgments}
We thank Alan Chamberlin for his help with setting up the infrastructure of the system and Paul Chodas, Ryan Park, Giovanni Valsecchi, and an anonymous reviewer for their comments about this manuscript. This research was carried out at the Jet Propulsion Laboratory, California Institute of Technology, under a contract with the National Aeronautics and Space Administration.

\appendix

\section{The $b$-plane\label{App:b-plane}}

The relative geometry of the encounter is conveniently characterized on the $b$-plane, which is defined as the plane normal to the asymptote directed along the incoming hyperbolic excess velocity vector ($\myvec{v}_{\infty}$) and centered at the Earth. The $b$-plane is defined using the osculating orbit of the asteroid at the time of the closest approach.

The $b$-plane is uniquely defined by the orthogonal frame $\{\myvec{u}_r, \myvec{u}_t, \myvec{u}_s\}$, where
\begin{equation}
    \myvec{u}_s = \frac{\myvec{v}_{\infty}}{||\myvec{v}_{\infty}||}, \qquad \myvec{u}_t = \frac{\myvec{k}\times\myvec{u}_s}{||\myvec{k}\times\myvec{u}_s||},\qquad \myvec{u}_r = \myvec{u}_s\times\myvec{u}_t.
\end{equation}
The reference unit vector $\myvec{k}$ follows the $-z$ direction in the equatorial ICRF reference frame \citep{farnocchia2019planetary}. The $b$-plane coordinates of the point where the asymptote pierces the $b$-plane are commonly denoted $\myvec{b}=[b_R, b_T]^\top$, with $b_R = \myvec{b}\cdot\myvec{u}_r$ and $b_T = \myvec{b}\cdot\myvec{u}_t$. The norm $||\myvec{b}||$ is called the impact parameter, which satisfies $||\myvec{b}||=h/v_\infty$ given the angular momentum $h$.

The condition for an impact with Earth is that the osculating periapsis radius be lower than one Earth radius. This condition yields the relation
\begin{equation}\label{Eq:bplane_cond}
    ||\myvec{b}|| \leq \lambda R_\earth, \qquad \lambda=\sqrt{1 + \frac{2\mu_\earth}{R_\earth v_\infty^2}},
\end{equation}
where $\mu_\earth$ is Earth's gravitational parameter. The scaling factor $\lambda$ accounts for the hyperbolic motion of the asteroid around Earth and it is a function of the $v_\infty$ of the encounter. Based on this definition, the scaled $b$-plane coordinates \citep{chodas1999predicting} take the form
\begin{equation}\label{Eq:components_b_scaled}
    {\myvec{b}_{s}} = \myvec{b}/\lambda = [{b}_{s,R},{b}_{s,T}]^\top.    
\end{equation}
The impact condition in Eq.~\eqref{Eq:bplane_cond} transforms into $||\myvec{b}_{s}||<R_\oplus$.

The construction of the $b$-plane presented in this section assumes that the osculating orbit of the asteroid relative to Earth is hyperbolic. In some cases, like asteroids temporarily captured by Earth's gravity, the orbit will not be hyperbolic so the $b$-plane will not be defined. The encounter is then characterized on the modified target plane \citep{milani1999asteroidII,yeomans1994predicting}, which is constructed similarly to the $b$-plane. However, the hyperbolic excess velocity is replaced with the asteroid's velocity relative to Earth at perigee. The impact pseudo-observation residuals are then defined in terms of the coordinates on the modified target plane, $(X,Y)$, while the rest of the algorithm remains unchanged. For a more detailed definition of the $b$-plane, the reader is referred to the work by \citet{hintz2015orbital} and \citet{farnocchia2019planetary}.

\end{document}